%% file: manuscript.tex
\documentclass[twocolumn, 3p, number, sort&compress]{elsarticle}

\usepackage[utf8]{inputenc}
\usepackage[T1]{fontenc}

\biboptions{sort&compress,super}

\usepackage{amsmath,amssymb}
\usepackage{graphicx}
\usepackage{balance}
\usepackage{hyperref}
\hypersetup{hidelinks}
\usepackage{csquotes}

\usepackage{mathrsfs} 
\usepackage{bm}
\usepackage{fancyvrb}
\usepackage{color, soul}
\usepackage[usenames,dvipsnames]{xcolor}

\usepackage{widetext} 

\usepackage{multirow}
\usepackage{pdflscape,array,booktabs}
\renewcommand{\arraystretch}{1}

\usepackage{siunitx}
\usepackage{listings}
\newcommand\subroutine[1]{\texttt{#1}}
\newcommand\module[1]{\texttt{#1}}
\newcommand\function[1]{\texttt{#1}}
\newcommand\keywd[1]{\texttt{#1}}
\newcommand\library[1]{\texttt{#1}}

\newcommand{\out}{magenta}

\newcommand{\tmatrices}{\emph{T}-matrices}
\newcommand{\tmatrix}{\emph{T}-matrix}
\newcommand{\FORTRAN}{\textrm{Fortran}}
\newcommand{\fortran}{\textrm{Fortran}}
\newcommand{\LAPACK}{\textsc{lapack}}
\newcommand{\lapack}{\textsc{lapack}}
\newcommand{\BLAS}{\textsc{blas}}
\newcommand{\blas}{\textsc{blas}}
\newcommand{\MSTM}{\textsc{mstm}}
\newcommand{\SMARTIES}{\textsc{smarties}}

\newcommand{\TERMS}{\textsc{terms}}
\newcommand{\terms}{\textsc{terms}}

\newcommand{\nmax}{n_\text{max}}
\newcommand{\lmax}{l_\text{max}}
\newcommand{\km}{k}

\newcommand{\bc}{\mathbf{c}}
\newcommand{\bP}{\mathbf{P}}
\newcommand{\rbc}{\widetilde{\mathbf{c}}}
\newcommand{\ba}{\mathbf{a}}
\newcommand{\rba}{\widetilde{\mathbf{a}}}

\newcommand{\rbe}{\widetilde{\mathbf{e}}}

\newcommand{\br}{\mathbf{r}}
\newcommand{\bd}{\mathbf{d}}
\newcommand{\bR}{\mathbf{R}}
\newcommand{\bS}{\mathbf{S}}
\newcommand{\bx}{\mathbf{x}}
\newcommand{\bE}{\mathbf{E}}
\newcommand{\rbE}{\widetilde{\mathbf{E}}}
\newcommand{\bM}{\mathbf{M}}
\newcommand{\bN}{\mathbf{N}}
\newcommand{\bW}{\mathbf{W}}

\newcommand{\bI}{\mathbf{I}}
\newcommand{\bZ}{\mathbf{Z}}
\newcommand{\bT}{\mathbf{T}}
\newcommand{\rbM}{\widetilde{\mathbf{M}}}
\newcommand{\rbN}{\widetilde{\mathbf{N}}}
\newcommand{\rbW}{\widetilde{\mathbf{W}}}
\newcommand{\rbw}{\widetilde{\mathbf{w}}}
\newcommand{\bD}{\mathbf{D}}
\newcommand{\rbD}{\widetilde{\mathbf{D}}}
\newcommand{\bO}{\mathbf{O}}
\newcommand{\rbO}{\widetilde{\mathbf{O}}}
\newcommand{\ldoc}{\mathscr{C}}
\newcommand{\ldocbar}{\overline{\mathscr{C}}}

\newcommand{\ldocoabar}{\langle\overline{\mathscr{C}}\rangle}
\newcommand{\trace}{\ensuremath{\operatorname{Tr}}} 
\DeclareMathOperator{\tr}{Tr}
\newcommand{\bn}{\mathbf{n}}

\newcommand{\bB}{\mathbf{B}}

\newcommand{\bhelB}{\mathbf{Z}}
\newcommand{\bhelBr}{\bhelB_{\scriptscriptstyle R}}
\newcommand{\bhelBl}{\bhelB_{\scriptscriptstyle L}}
\newcommand{\bheltBr}{\widetilde{\bhelB}_{\scriptscriptstyle R}}
\newcommand{\bheltBl}{\widetilde{\bhelB}_{\scriptscriptstyle L}}

\newcommand{\tinyL}{\scriptscriptstyle L}
\newcommand{\tinyLL}{\scriptscriptstyle LL}
\newcommand{\tinyR}{\scriptscriptstyle R}
\newcommand{\tinyRR}{\scriptscriptstyle RR}
\newcommand{\tinyLR}{\scriptscriptstyle LR}
\newcommand{\tinyRL}{\scriptscriptstyle RL}

\newcommand{\bUl}{\mathbf{U}^{\scriptscriptstyle (L)}}
\newcommand{\bUr}{\mathbf{U}^{\scriptscriptstyle (R)}}
\newcommand{\bVl}{\mathbf{V}^{\scriptscriptstyle (L)}}
\newcommand{\bVr}{\mathbf{V}^{\scriptscriptstyle (R)}}
\newcommand{\bUhl}{\mathbf{U}^{\dagger\scriptscriptstyle (L)}}
\newcommand{\bUhr}{\mathbf{U}^{\dagger\scriptscriptstyle (R)}}
\newcommand{\bVhl}{\mathbf{V}^{\dagger\scriptscriptstyle (L)}}
\newcommand{\bVhr}{\mathbf{V}^{\dagger\scriptscriptstyle (R)}}

\title{Multiple scattering of light in nanoparticle assemblies:\\%
user guide for the \TERMS\ program}
\hypersetup{draft}
\begin{document}
\author{D.~Schebarchov}
\ead{dmitri.schebarchov@gmail.com}
\author{A.~Fazel-Najafabadi}
\ead{atefeh.fazelnajafabadi@vuw.ac.nz}
\author{E. C.~Le~Ru}
\ead{eric.leru@vuw.ac.nz}
\author{B.~Auguié\corref{cor1}}
\ead{baptiste.auguie@vuw.ac.nz}
\address{The MacDiarmid Institute for
  Advanced Materials and Nanotechnology \\School of Chemical and
  Physical Sciences, Victoria University of Wellington,\\ PO Box 600,
  Wellington 6140, New Zealand}
\cortext[cor1]{Corresponding authors}

\onecolumn
\begin{abstract}
We introduce \TERMS, an open-source Fortran program to simulate near-field and far-field optical properties of clusters of particles. The program solves rigorously the Maxwell equations via the superposition \tmatrix\ method, where incident and scattered fields are decomposed into series of vector spherical waves. 

\TERMS\  implements several algorithms to solve the coupled system of multiple scattering equations that describes the electromagnetic interaction between neighbouring scatterers. From this formal solution, the program can compute a number of physically-relevant optical properties, such as far-field cross-sections for extinction, absorption, scattering and their corresponding circular dichroism, as well as local field intensities and degree of optical chirality. By describing the incident and scattered fields in a basis of spherical waves the \tmatrix\ framework lends itself to analytical formulas for orientation-averaged quantities, corresponding to systems of particles in random orientation; \TERMS\ offers such computations for both far-field and near-field quantities of interest. This user guide introduces the program, summarises the relevant theory, and is supplemented by a comprehensive suite of stand-alone examples in the website accompanying the code.
\end{abstract}
\maketitle
\setcounter{tocdepth}{2}
\tableofcontents

\twocolumn
\section{Introduction}

\TERMS\ -- acronym for \tmatrix\ for Electromagnetic Radiation with Multiple Scatterers -- is a suite of \FORTRAN\ 90 routines to simulate light scattering by rigid clusters of particles immersed in a homogeneous, non-absorbing medium. The calculation is based on the superposition \tmatrix\ (STM) method, an extension of Waterman's \tmatrix\ formalism\cite{waterman1965matrix, waterman1969new,waterman1971symmetry,MishchenkoTL02} to multiple scatterers\cite{PetersonS73,Chew:1990aa, Mackowski91}. The incident and scattered fields are expanded into series of vector spherical wave functions (VSWFs), which can be interpreted as a multipolar decomposition.
For linear media, the coefficients describing the scattered field follow a linear relationship with those of the known incident field; this linear relationship is expressed through the so-called \tmatrix, which encodes the full information about a scatterer's linear optical properties and its response to an arbitrary incident excitation. Where several particles are present, light scattered by one particle can contribute to the excitation of the others; the self-consistent set of exciting and scattered fields from each particle, and the cluster as a whole, is expressed in the STM framework as a linear system of equations involving the single particle \tmatrices, and translation matrices to transform the VSWFs from one particle to another. The solution of this system of equations enables the calculation of near-field quantities as well as far-field cross-sections, for specific directions of incidence and polarisation, or after analytical orientation-averaging. 

In principle many types of particle shapes can be used in \TERMS, provided an external program can calculate and export their corresponding \tmatrix. \TERMS\ provides built-in calculations of single-particle \tmatrices\ for homogeneous and multi-layered spheres, and our Matlab code \SMARTIES\ can export accurate \tmatrices\ for oblate and prolate spheroidal particles in a compatible format\cite{Somerville:2016aa}. The maximum number of particles that \TERMS\ can consider is typically about a few hundred for standard computers and small maximum multipolar order, although larger systems could be modelled using an iterative linear solver\cite{Mackowski:2012ug,Yurkin:2016ta,Markkanen:2017aa} or implementing a hierarchical fast multipole method\cite{Solis:2014uo,Markkanen:2017aa}.

This guide aims to describe the program from a user's perspective, illustrate the types of calculations that it can perform, and highlight its strengths relative to other computational methods. The code is released as open-source, and we welcome contributions from the community. A dedicated website\cite{Schebarchov:2021ut} provides stand-alone examples to illustrate the program's capabilities in specific applications. 
\subsection{General features}
From a generic description of the scattering problem, consisting in the position and orientation of $N$ particles, dielectric functions or input \tmatrix\ for each particle, and the incident wavelength(s), the program can perform three main types of simulations:
\begin{enumerate}
\item \emph{Near-field mode}, to map local fields and derived quantities at fixed incidence, or with orientation-averaging.
\item \emph{Far-field mode}, to calculate cross-sections (extinction, scattering, and absorption, as well as corresponding linear and circular dichroism) at fixed incidence and with orientation-averaging.
\item \emph{Polarimetry mode}, to calculate Mueller matrices, Stokes parameters, and differential scattering cross-sections at specified scattering angles.
\end{enumerate}
At runtime, the program sets up a linear system of equations in the form $\mathbf{Ax}=\mathbf{b}$, where the matrix $\mathbf{A}$ is constructed from a given set of single-particle \tmatrices, particle coordinates and orientations, and the vector (or matrix) $\mathbf{b}$ characterises the specified incident plane wave excitation(s). The unknown $\mathbf{x}$ determines the self-consistent field exciting each scatterer, as described in more details in Sec.~\ref{sec:principles}. The linear system is then solved using one of several schemes selected by the user:
\begin{itemize}
\item[0.] Application of a (direct) solver to determine $\mathbf{x}$, corresponding to the particle-centred scattering coefficients for one or more specific incident field(s), $\mathbf{b}$.
\item[1.] Direct inversion of the matrix $\mathbf{A}$ to determine the particle-centred \tmatrices\ for the cluster of particles.\cite{StoutAL02}
\item[2.] Stout~\emph{et al.}'s\,\cite{StoutAL02,StoutAD08} iterative scheme for calculating the particle-centred \tmatrices.
\item[3.] Mackowski \& Mishchenko's\,\cite{Mackowski91,Mackowski94,Mackowski96,MackowskiM11,Mackowski:2012ug} scheme for calculating the particle-centred \tmatrices.
\end{itemize}
Implementation of these multiple solution schemes in a modular code-base is a core feature in \TERMS; we hope it will prove useful for designing, testing, and benchmarking various methods, and perhaps lead to the implementation of new improved algorithms.

The most notable features of \TERMS\ include:
\begin{itemize}
\item Export of the collective \tmatrix\ describing the entire cluster of particles.
\item Import of general \tmatrices, which can be pre-generated using \TERMS\ or another program, such as \SMARTIES\ (for spheroids)\cite{Somerville:2016aa}.
\item Built-in calculation of individual \tmatrices\ for stratified/coated spheres described by Mie theory\cite{LeRuE08}.
\item Calculation of partial absorption cross-sections in each layer of coated spheres, following Mackowski\cite{Mackowski:1990aa}.
\item Calculation of orientation-averaged far-field cross-sections and associated circular dichroism\cite{Suryadharma18}.
\item Calculation of orientation-averaged near-fields\cite{StoutAD08} and optical chirality\cite{fazel2021orientation}.
\item Calculation of the Mueller matrix and Stokes parameters for specific incidence and scattering angles\cite{MishchenkoTL02}.
\item Possible compilation with all double-precision variables promoted to quad-precision\cite{schebarchov2019mind}.
\item Export the output results in plain text or "HDF5" file format\cite{hdf5}.
\end{itemize}
\subsection{Relation to other codes}
\TERMS\ belongs to the family of codes implementing the superposition \tmatrix\ method for collections of scatterers. Other implementations have been described in the literature \cite{Mackowski91, Garcia-de-Abajo:1999aa,Vargas:93,xu1995electromagnetic,StoutAL01,StoutAL02,StoutAD08,Suryadharma18,Fruhnert:2017aa} (for a comprehensive review, we refer the reader to Ref.~\citenum{Mishchenko:2020ul}); available open-source programs include that of Mishchenko \& Mackowski for spherical particles and optically-active media (\MSTM) \cite{MackowskiM11}, and for nonspherical particles the recent additions of \textsc{FastMM} by Markkanen and Yuffa\cite{Markkanen:2017aa} and \textsc{qpms} by Ne\v{c}ada and T\"{o}rm\"{a}\cite{Torma:2021vz}.

Among the many available techniques to solve light scattering problems\cite{Kahnert:2003wb}, the STM method holds distinct advantages over purely numerical techniques such as the Finite Elements Method (FEM)\cite{Jin:2014wn}, the Discrete Dipole Approximation (DDA)\cite{Yurkin:2007wv}, or the Finite Differences Time Domain (FDTD) method\cite{Archambeault:1998tg}. Unlike STM, these techniques require discretising the whole cluster geometry and solving the full electromagnetic problem for every direction of incidence. Other notable advantages include:
\begin{itemize}
\item Orientation-averaged far-field properties can be obtained at very little computational cost, with analytical formulae\cite{mishchenko1989interstellar,mishchenko1990extinction,Khlebtsov92, borghese2007scattering,Suryadharma18}. Orientation-averaged near-field quantities can also be computed\cite{StoutAD08,fazel2021orientation}, albeit with some computational overhead, providing analytical benchmark results\cite{Fazel-Najafabadi:2021ud}.
\item For clusters of several identical particles only one \tmatrix\ needs to be calculated.
\item Within its domain of validity the Extended Boundary Condition Method (EBCM), and the \tmatrix\ framework more broadly, is typically faster and more accurate than competing methods, and is therefore often used for benchmark calculations\cite{MishchenkoTL02}.
\item The multipolar decomposition of electromagnetic fields can provide physical insight into complex optical responses\cite{Mun:2020ww}.
\end{itemize}
It should be noted that the STM method is not without its limitations,
\begin{itemize}
\item Closely-spaced scatterers can lead to inaccurate results, or require very large multipolar orders, and the exact domain of applicability of the method in such situations is not fully-understood\cite{schebarchov2019mind}. Some proposals to overcome this issue have recently been demonstrated\cite{Theobald:2017ty}, and may be implemented in \TERMS\ in the future.
\item The calculation of local fields in the vicinity of elongated nanoparticles is limited by the Rayleigh Hypothesis\cite{auguie_numerical_2016}. 
\item Our particular implementation is limited to relatively small numbers of particles (a few tens to hundreds, on a typical workstation, and depending on their size parameter).
\item Numerical instabilities arise at high maximum multipolar order (from approximately $\nmax\approx 30$ typically), preventing the calculation of accurate \tmatrix\ elements in double-precision, and leading to ill-conditioning of matrices.
\item Nonspherical particle shapes require first computing the \tmatrix\ with an external program. \TERMS\ has built-in functions for homogeneous and multi-layered spheres, and for non-spherical particles the \tmatrix\ can be obtained from a variety of methods, from Mishchenko's EBCM implementation for axisymmetric particles\cite{MishchenkoTL02} and \SMARTIES\cite{Somerville:2016aa} in particular for spheroids, the surface-integral equation (SIE) method\cite{Homer-Reid:2013aa}, the volume-integral equation method\cite{Markkanen:2017aa}, and other algorithms have been proposed for the Discrete Dipole Approximation\cite{Loke:2009aa}, or general solvers such as the Finite-Element Method\cite{Fruhnert:2017aa}. 
\end{itemize}
Different superposition \tmatrix\ algorithms have been proposed, with their own strengths and weaknesses depending on the type of particles and their configuration; an important feature of \TERMS\ is that it is possible to compare several algorithms and choose the most suitable for a given problem. For complex geometries, especially compact ones, the invariant-embedding \tmatrix\ method\cite{Johnson:1988vd}, the Surface Integral Equation\cite{Kern:2009tj} or Volume Integral Equation methods\cite{Homer-Reid:2013aa} may provide better alternatives. Our implementation also does not currently consider periodic arrays of scatterers\cite{Stefanou:1992uk,Fruhnert:2017aa,Torma:2021vz}, or the presence of a substrate\cite{Doicu:2006ua,Garcia-de-Abajo:2007wz}.
\subsection{Aims of this manual}
\TERMS\ is accompanied by a comprehensive set of examples available online\cite{Schebarchov:2021ut}; this user guide aims to provide a useful complement introducing i) the necessary background information about the method; ii) the initial steps required to install and run the program; iii) a high-level description of the program and its capabilities.
\subsection{Licensing}
\TERMS\ is made available under the Mozilla Public License Version 2.0, but note that parts of the code include external \FORTRAN\ libraries under different licencing, such as \LAPACK\ (BSD), and HDF5 routines (copyright The HDF Group)\cite{hdf5}. 
\subsection{Disclaimer and request for feedback}

The \TERMS\ program is provided "as is", without warranty of any kind. While we have tested the program in a large number of configurations we cannot provide any guarantee as to the accuracy or validity of simulation results obtained with the program. The user is strongly encouraged to perform their own reference checks against other methods, but also internal consistency checks by switching the solution method, increasing the multipolar order, and if necessary using quad precision.
%

We welcome comments, reports of errors, and suggestions of new features, which can be addressed directly to the authors or via the code's hosting website.

\section{Getting started}

%
\begin{figure*}[!htpb]
\center
\includegraphics[width=1\textwidth]{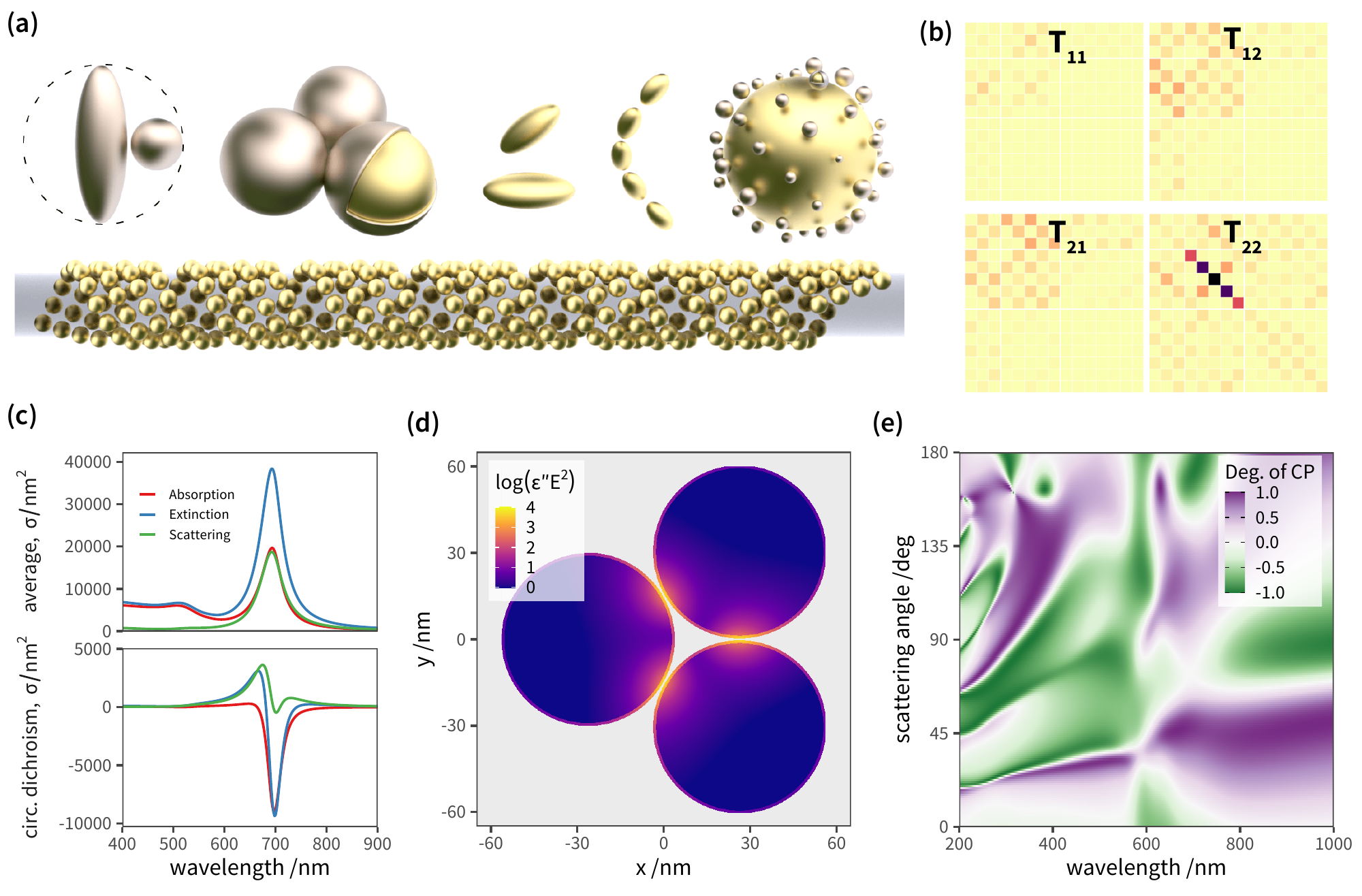}
\caption{Illustrative overview of \TERMS. (a) Pictorial representations of nanoparticle clusters studied with \terms\ (Left to right: closely-spaced dimer\cite{schebarchov2019mind}, trimer of Au@Pd core-shell antennas\cite{Lee:2020aa}, chiral dimer and helix of Au spheroids\cite{Fazel-Najafabadi:2021ud} (also bottom helix), hybrid antenna-satellite photocatalyst\cite{Herran:2021wh}). (b) Collective \tmatrix\ of a chiral dimer of prolate spheroids (after online example 13\cite{Schebarchov:2021ut}; the colour maps the modulus of the \tmatrix\ elements, here truncated at $\nmax=3$). (c) Far-field spectra of orientation-averaged cross-sections (absorption, scattering, extinction, and their corresponding circular dichroism in the bottom panel); the structure consists of a chiral dimer of prolate Au spheroids in water\cite{Fazel-Najafabadi:2021ud}. (d) Near-field map of $\Im(\varepsilon) |\mathbf{E}|^2$ in a trimer of Au@Pd core-shell antennas\cite{Lee:2020aa}. (e) Dispersion map of the degree of circular polarisation displayed by a helix of five prolate Au spheroids (after online example 08\cite{Schebarchov:2021ut}).}
\label{fig:overview}
\end{figure*}
Figure~\ref{fig:overview} displays a partial overview of \TERMS' capabilities, with calculation results taken from the online documentation\cite{Schebarchov:2021ut}, which includes over 20 self-contained examples illustrating all the different options for using \TERMS. We do not repeat these examples in this user guide but instead provide the basic common starting point which can be adapted for any specific use case.
\subsection{Installation}
The code was developed and tested predominantly on standard personal desktop and laptop computers running Linux (Ubuntu 18.04 LTS) and MacOS, as well as the R\=apoi HPC Cluster at Victoria University of Wellington. We've also successfully installed and run \terms\ on Windows via the Windows Subsystem for Linux (WSL\,2). Our Linux configuration includes: the \texttt{gfortran} compiler in \texttt{gcc} version 7.4.0, \texttt{HDF5} software with libraries "\library{libhdf5-dev}", \blas\ "\library{libblas-dev}" and \lapack\ "\library{liblapack-dev}". We advise using a fairly recent \FORTRAN\ 90 compiler (\texttt{gcc} versions below 6 have caused problems), and recent HDF5 release "\library{HDF5-1.12.1}"\cite{hdf5}.

There are two ways for producing the executable file:
\begin{itemize}

\item (Recommended) using \library{Cmake}, with parameters defined in \texttt{CMakeLists.txt}:
\begin{verbatim}
> cd build
> cmake ..
> make
\end{verbatim}
will produce an executable \texttt{terms} for your machine, which you can leave in its location or move elsewhere.\\[1em]Alternatively,
\item A basic script is provided under \texttt{build/buildTERMS.sh} to specify the compilation options (double vs quad precision, debug mode, and use of a system's \lapack). 
\begin{verbatim}
> cd build
> bash buildTERMS.sh
\end{verbatim}
will produce an executable \texttt{terms} for your machine, which you can leave in its location or elsewhere. 
%
%

%
\end{itemize}

The executable reads user-defined instructions describing the scattering problem from an input file, and is called as follows:
\begin{verbatim}
> ./terms inputfile > messages.log 
\end{verbatim}

The results of calculations are stored in specific output files in the current directory and displayed in the terminal together with any errors and warnings (it can be convenient to redirect the standard output to a log file, as in the example above).
\subsection{Initial steps}
The main input parameters are read from a plain text input file (line by line and from left to right; blank lines are ignored). Each line is interpreted as a sentence and split into space-separated words. The first (left-most) word is interpreted as a case-sensitive keyword, and the subsequent words as arguments for that keyword. In each sentence, text from the first word starting with the hash character (\#) is interpreted as a human-readable comment and thus ignored by the program. All the supported keywords and corresponding arguments are documented in Appendix \ref{appendix:keywords}. The order of keywords generally doesn’t matter, with just two exceptions: \keywd{ModeAndScheme} must be the first keyword, and \keywd{Scatterers} must be the last.
\subsection{Minimal example}
We first illustrate the use of \TERMS\ on a simple case, the calculation of
far-field spectra for absorption, scattering, and extinction with a
structure consisting of four gold spheres immersed in water. Most simulation
parameters are kept to their default values.

This simulation uses the following \texttt{input} file,
\begin{verbatim}
ModeAndScheme 2 3
Wavelength 300 900 300
Medium 1.7689  # epsilon of water

Scatterers  4 
Au  31.5     0       -50  30               
Au -31.5     0       -50  30               
Au  22.2738  22.2738  50  30             
Au -22.2738 -22.2738  50  30               
\end{verbatim}
The program is run with the command
\verb+./terms input > log+, where the \texttt{log} file contains information about the simulation (how detailed depends on the optional \keywd{Verbosity} argument). The output for this simulation consists of a number of plain text files, storing the far-field cross-sections:
\begin{itemize}
\item
  Files \texttt{cs(Abs|Ext|Sca)OA} contain
  orientation-averaged cross-sections.
\item
  Files \texttt{cd(Abs|Ext|Sca)OA} contain
  orientation-averaged optical activity.
\item
  Files
  \texttt{cs(Abs|Ext|Sca)(1X|2Y|3R|4L)}
  contain fixed-orientation cross-sections for the respective
  polarisation ('\texttt{X}', '\texttt{Y}': 2 orthogonal linear polarisations; '\texttt{R}', '\texttt{L}': right and left circular polarisations).
\item
  Files
  \texttt{csAbs(1X|..)\_scat(00i)coat(j)}
  contain partial absorption cross-sections inside multi-layered spheres.
\end{itemize}
This plain text output can become inconvenient when running many simulations; \terms\ provides an option to produce a single Hierarchical Data Format (\texttt{HDF5}) output file\cite{hdf5}, with the output quantities stored under individual fields instead of separate files. The \texttt{HDF5} file format can be read in many other programs, using e.g the built-in \texttt{h5read} function in Matlab, or packages \texttt{rhdf5} for R, \texttt{h5py} for Python, \texttt{HDF5.jl} for Julia, to list only a few popular options.

The documentation's website features many minimal examples of use for each option of the program, and with various cluster configurations.

\subsection{Range of validity}
The \tmatrix\ method is often used as a benchmark for other numerical techniques such as DDA or FEM, as it provides very accurate results. A sufficiently-high value of the maximum multipole order, $\nmax$, should be chosen for each simulation, and convergence of the results with increasing $\nmax$ is often a good indicator of the accuracy of the results. \TERMS\ performs internal checks of convergence for the far-field cross-sections, by comparing the relative error between successive partial sums over multipole orders 1 to $\nmax$. We strongly advise users to monitor the messages and check for issues with convergence. It is also useful to re-run calculations with a higher value of $\nmax$ and check that the results do not differ. 
In near-field calculations a higher $\nmax$ value is generally needed, and we find that values above 30 can require switching to quad precision. The challenging case of nonspherical particles with strongly-overlapping circumscribed spheres pushed some calculations to use $\nmax$ above 50; even with quad precision arithmetic the accuracy eventually deteriorates (above 60, typically). We emphasise that these are extreme cases; in many standard situations a low value of $\nmax$ is sufficient (8 is the default value). The coupled-dipole method, widely used in nano-optics, corresponds roughly to setting $\nmax=1$.

Single-particle \tmatrices\ computed with Mie theory are generally accurate up to $\nmax=60$. Following Wiscombe's criterion, this corresponds to a size parameter of 45, or a sphere radius of 2 microns in vacuum for visible light. For spheroids, \SMARTIES\ enables accurate calculation of \tmatrix\ elements with an aspect ratio of up to 100, and similar size limitations as Mie theory\cite{Somerville:2016aa}. 

Multiple-scattering generally introduces a loss of precision compared to single-particle calculations, and requires larger values of $\nmax$. The user is advised to consider the different solution schemes implemented in \TERMS, as they can offer substantial benefits in specific situations. For instance, Stout and co-workers introduced a balancing scheme\cite{StoutAD08} that stabilises the numerical calculations and proves very effective for closely-spaced resonant particles. \TERMS\ has extended this improvement to other schemes by default (controlled with the keyword \keywd{StoutBalancing}). A dramatic difference between Scheme 2 and 3 is observed when particles are widely-separated: our implementation of Mackowski \& Mishchenko's scheme fails where separations are above a few hundred nanometres even at large $\nmax$, while Stout's scheme maintains good accuracy without requiring a $\nmax$ value much larger than dictated by the single-particle response. The key difference between the two schemes is that Mackowski \& Mishchenko's translates all VSWFs to a common origin, while Stout's maintains particle-centred expansions throughout\cite{Mackowski94, Mackowski96, MackowskiM11, StoutAD08, StoutAL02, auger2008local}.

The performance of Mackowski \& Mishchenko's scheme is usually very good, in both accuracy and speed, and is chosen as the default. 

The results of \TERMS\ calculations have been validated against Mackowski \& Mishchenko's \MSTM\ code for collections of spheres\cite{MackowskiM11}, and against a commercial Finite Element package (Comsol\cite{Multiphysics:1998tl}) for dimers of spheroids\cite{schebarchov2019mind}. 

\section{Underlying principles of the code}
\label{sec:principles}

\input{theory.tex}

\section{Conclusion and outlook}
\label{sec:outlook}

We have introduced \terms, an open-source \fortran\ program to simulate light scattering by rigid clusters of nanoparticles, in fixed or random orientation with respect to incident light. \terms\ implements several superposition \tmatrix\ algorithms and recently-derived formulas for analytical orientation-averaging of far-field and near-field optical properties. This manuscript provides a brief summary of the method and references the key formulas implemented in the program. A companion website\cite{Schebarchov:2021ut} includes a comprehensive suite of self-contained examples illustrating the program's capabilities in realistic calculations. 
We hope this program will be useful to the light scattering community of researchers, and we welcome contributions to extend the program's use cases. As noted in the introduction, the superposition \tmatrix\ method has been implemented in several other publicly-available programs, each with its own set of features, and we welcome collaboration to combine these efforts. We conclude below with an outlook of the possible extensions we are considering for the future development of \terms.\\

\noindent\emph{Code improvements}
\begin{itemize}
\item Optional use of an iterative solver to solve large linear systems
\item Import/export of \tmatrices\ in HDF5 format
\item Import of \tmatrices\ for more general particle shapes, from Scuff-EM\cite{Homer-Reid:2013aa}
\item Built-in calculation of spheroid \tmatrices\ (port of \SMARTIES\cite{Somerville:2016aa})
\item Improved methods for the calculation of TACs\cite{Chew:2008vg}
\item Additional built-in material dielectric functions
\item Optimisation of time-consuming calculations
\end{itemize}

\noindent\emph{New features}
\begin{itemize}
\item Calculation of internal fields for non-spherical particles obtained via EBCM\cite{Somerville:2016aa} (exporting matrix $\mathbf{R}=\mathbf{Q}^{-1}$)
\item Orientation-averaged internal fields for coated spheres and nonspherical particles, adapting Ref.~\citenum{StoutAL02}
\item Orientation-averaged partial absorptions for layered particles
\item Orientation-averaged absorption and scattering circular dichroism in
  Stout's scatterer-centred formalism (not from the collective \tmatrix)
\item Chiral media, following Ref.~\citenum{Mackowski:wc}
\item \tmatrix\ for anisotropic core-shell spheres, following Ref.~\citenum{Tang:2021ux}
\item \tmatrix\ for coated spheroids, following Refs.~\citenum{Peterson:1974te,Quirantes:2005vb}
\item Dipolar incident field
\item \tmatrix\ of model molecules, following Ref.~\citenum{Fernandez-Corbaton:2020wh}
\item Conversion from \tmatrix\ to ``Higher-Order Polarizability Tensors'', following Ref.~\citenum{Mun:2020ww}
\item Extension to infinite periodic arrays, following Ref.~\citenum{Torma:2021vz}
\item Integral representation of near-fields in the Rayleigh region, following Ref.~\citenum{auguie_numerical_2016}
\item Plane-wave expansion for particles with intersecting circumscribed spheres, following Ref.~\citenum{Theobald:2017ty}
\item Geometry optimisation (via external libraries)
\end{itemize}


%
We also consider porting the codebase to the Julia language\cite{Julia-2017}, to benefit from an interactive environment to develop and test new features, without sacrificing run-time performance.

\section*{Acknowledgments}
The authors acknowledge the support of the Royal Society Te Ap\=arangi for support through a Rutherford Discovery Fellowship (B.A., grant number RDF-VUW1603), and the MacDiarmid Institute.
We dedicate this work to the memory of Michael I. Mishchenko, who had a tremendous influence on the electromagnetic and light scattering community, and left us too soon.

\bibliographystyle{elsarticle-num}
\bibliography{references}

\appendix
\renewcommand*{\thesection}{\Alph{section}}
\onecolumn
\section{Appendix}
\subsection{Keywords}
\label{appendix:keywords}
List of case-sensitive keywords and corresponding arguments supported
by \TERMS. Optional arguments are enclosed in square brackets (nested in
some cases).
%
\input{keywords.tex}

\clearpage
\subsection{Organisation of the code}
\label{sec:organisation}
%

%
\begin{figure*}[!htpb]
\center
\includegraphics[width=0.9\textwidth]{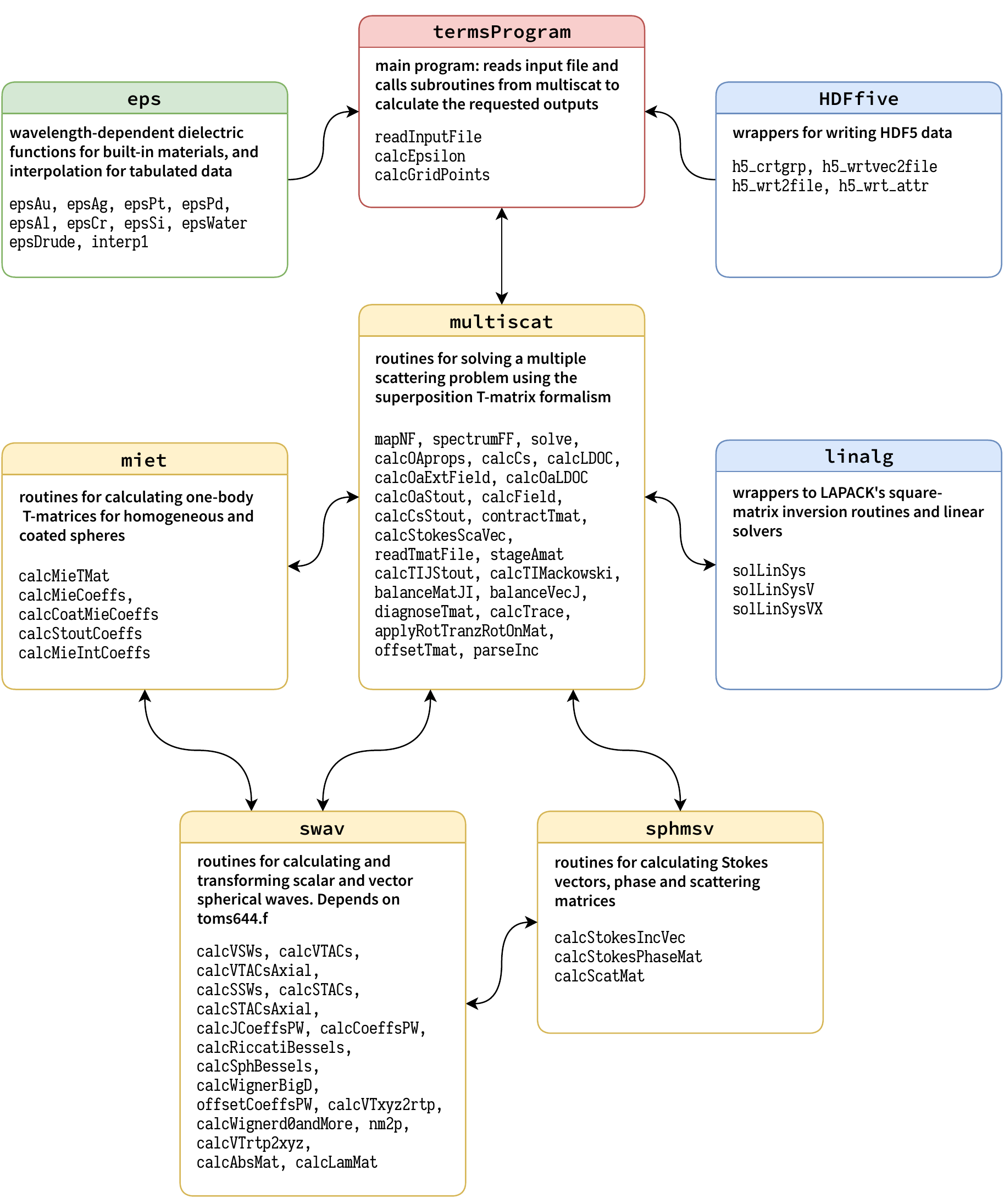}
\caption{Organigram of the code structure in \TERMS, organised in 8 modules. The \module{termsProgram} module is the general entry point to the program, dispatching the calculations to \module{multiscat}. In turn, \module{multiscat} uses subroutines from \module{miet}, \module{swav}, \module{sphmsv}, as well as linear algebra wrappers for \lapack\ in \module{linalg}. The module \module{eps} provides definitions of common dielectric functions and associated routines, while \module{HDFfive} provides wrappers for the HDF5 output data format.}
\label{fig:modules}
\end{figure*}

\subsection{List of subroutines}
\label{appendix:subroutines}
%
\input{subroutines.tex}
%
%
\end{document}

%% file: theory.tex

In the following presentation, the complex electric field is denoted by $\bE(\br,t)$, where $\br$ is a point coordinate and $t$ is time; we assume harmonic time dependence at angular frequency $\omega$, so that $e^{-i\omega t}$ factors out and is omitted from the rest of the discussion. 
\subsection{Vector spherical wave functions}
We define the vector spherical wave functions (VSWFs) as,
\begin{align}
\bM_{nm}^{(\zeta)}(k\br) = & \frac{1}{\sqrt{n(n+1)}} \nabla \times (\psi_{nm}^{(\zeta)} (k\br)\br), \\
\bN_{nm}^{(\zeta)}(k\br) = & \frac{1}{k} \nabla \times \bM_{nm}^{(\zeta)}(k\br). 
\end{align}
with $\km$ the wavenumber and 
\begin{equation} \label{eqn:psi_nm}
\psi_{nm}^{(\zeta)}(k\br) = z_{n}^{(\zeta)}(kr) Y_{nm}(\theta,\varphi),
\end{equation}
where $z_{n}^{(\zeta)}$ are spherical Bessel functions. For our purposes we only require $\zeta = 1$ ($z_{n}^{(1)} = j_{n}$, spherical Bessel functions of the first kind) and $\zeta = 3$ ($z_{n}^{(3)} = h_{n}$, spherical Hankel functions of the first kind), referred to as \emph{regular} and the \emph{irregular} functions, respectively, which are linearly independent. Henceforth, for brevity and notational convenience we refer to $\psi_{nm}^{(3)}$ as simply $\psi_{nm}$, and $\psi_{nm}^{(1)}$ as $\widetilde{\psi}_{nm}$. Furthermore, the tilde will also be placed over the coefficients (e.g.~$\rba$) to explicitly indicate a regular basis set. 

The spherical harmonics $Y_{nm}$ for $|m| \leq n$ we write as,
\begin{equation}
Y_{nm}(\theta,\varphi) = \gamma_{nm}\sqrt{n(n+1)} P_{n}^{m}(\cos\theta)\mathrm{e}^{\mathrm{i}m\varphi},
\end{equation}
where the associated Legendre functions $P_{n}^{m}(\cos\theta)$ are defined using the Condon-Shortley phase and
\begin{equation}
\gamma_{nm} := \sqrt{\frac{(2n+1)}{4\pi n(n+1)}\frac{(n-m)!}{(n+m)!}}.
\end{equation}
This convention is consistent\footnote{However, note that Mishchenko \emph{et al.}\cite{MishchenkoTL02} define their $\psi_{nm}^{(\zeta)}(k\br)$ as $z_{n}^{(\zeta)}(kr) P_{n}^{m}(\cos\theta)\mathrm{e}^{\mathrm{i}m\varphi}$, which must be multiplied by $\gamma_{nm}\sqrt{n(n+1)}$ to match our $\psi_{nm}^{(\zeta)}(k\br)$ in (\ref{eqn:psi_nm}).} with our main references.\cite{MishchenkoTL02, StoutAL02, LeRuE08, Chew92} 

Formally, $n$ can run from $0$ up to $\infty$, though numerically all series of VSWFs are truncated to some maximum multipole order $\nmax$. We also introduce the composite index $p(n,m)$ for convenience, defined as
\begin{equation}
p:=n(n+1)+m
\end{equation}
with,
\begin{align}
n  = & \mathrm{Int}(\sqrt{p}) \\ 
m  = & p-n(n+1).
\end{align}
A general regular solution to the Helmholtz equation can be expressed in the VSWF basis as,
\begin{align}
\rbE( k{\bf r}) & =  \sum\limits_{n=1}^{\nmax} \sum\limits_{m=-n}^{n} [\widetilde{a}_{1,nm}\widetilde{\bf M}_{nm}(k{\bf r}) + \widetilde{a}_{2,nm}\widetilde{\bf N}_{nm}(k{\bf r})] \nonumber \\
{}&= \sum_{s=1}^{2}\sum_{p=1}^{p_\mathrm{max}} \widetilde{a}_{s,p} \rbw_{s,p}(k\br)  \nonumber \\
{}&= \sum_{l=1}^{\lmax} \rbw_{l}(k\br) \widetilde{a}_{l} ~ =: ~ \rbW(k\br) \rba, \label{eqn:VectorGenSol}
\end{align}
where $\rba \in \mathbb{C}^{\lmax}$ is a column vector of coefficients, $\rbW = [\rbw_{1}, \rbw_{2},\dots,\rbw_{\lmax}]$ is a basis-set \emph{pseudo-matrix} of dimension $3\times \lmax$, i.e. a row vector composed of column vectors
\begin{equation}
\rbw_{l(s,n,m)} := \left\{
\begin{array}{lll}
\rbM_{nm} & \mathrm{for} & s = 1 \\
\rbN_{nm} & \mathrm{for} & s = 2
\end{array} \right.
\end{equation}
and $\lmax$ is the maximal value of another composite index $l$, introduced for convenience
\begin{align}
l & := (s-1)\nmax(\nmax + 2)\label{eqn:defLindex}  \\
 {} & \phantom{=} + n(n+1) + m \nonumber \\
& =   (s-1)p_\mathrm{max} + p \nonumber \\
 & \leq \lmax \nonumber\\
 \lmax & =2\nmax(\nmax + 2) = 2p_\mathrm{max}, \label{eqn:defLmax}
\end{align} 
where $s \in \{1,2\}$ is sometimes referred to as the \emph{parity} or \emph{mode} index, corresponding to either $\mathbf{M}$ or $\mathbf{N}$ functions.

The irregular variant of (\ref{eqn:VectorGenSol}) is obtained by simply removing the overhead tildes (e.g.~$\rbW \to \bW$), which corresponds to switching the radial dependence from $j_{n}(kr)$ to $h_{n}(kr)$ throughout. 
\subsection{The \tmatrix\ ansatz}
\label{sec:tmatrix}
\begin{figure}
\center
\includegraphics[width=0.5\textwidth]{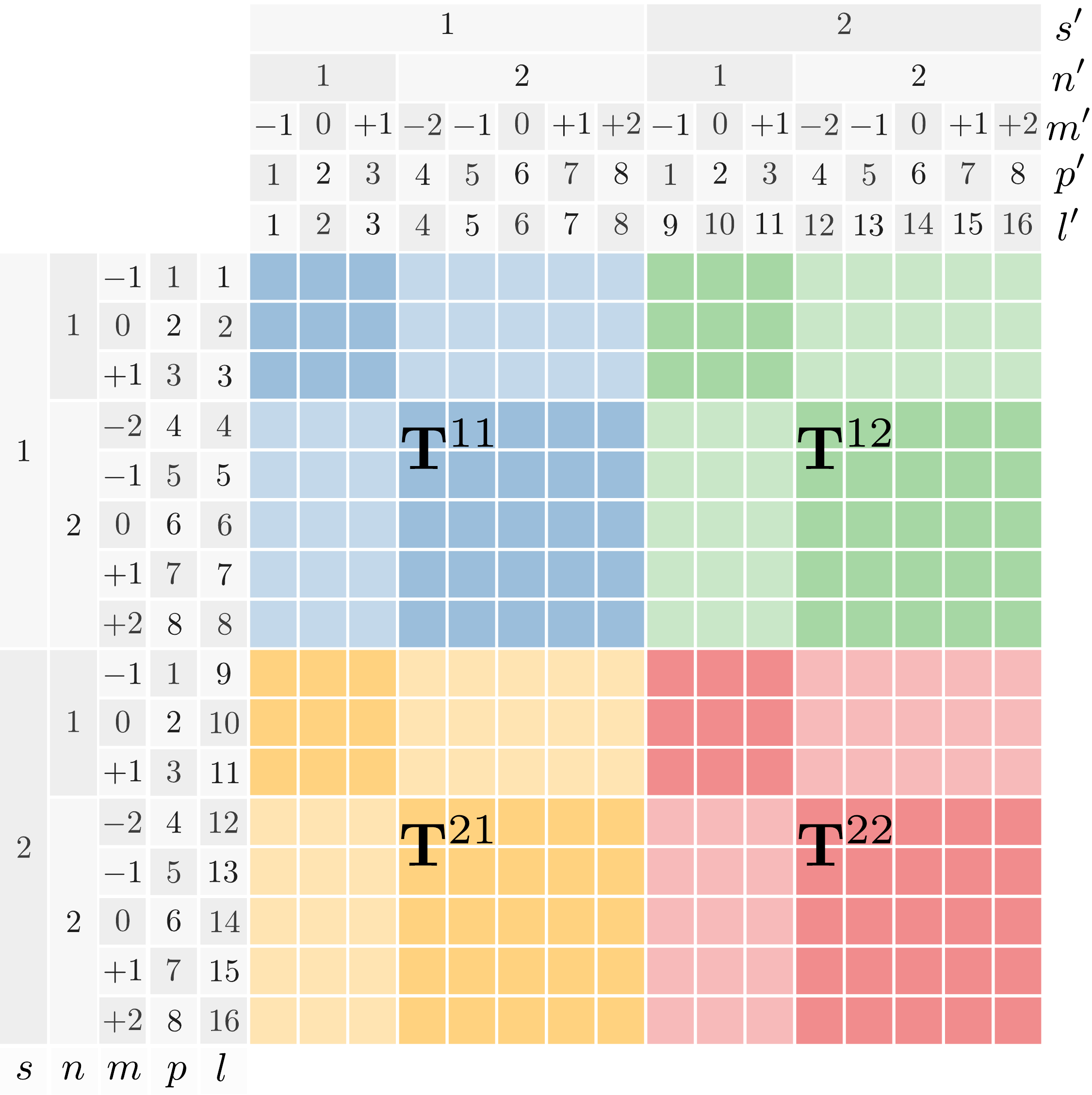}
\caption{Pictorial representation of a \tmatrix\ and relevant indices for $n_{max} = 2$. The matrix elements coupling $\bM_{nm}$ with $\bM_{n'm'}$ (magnetic--magnetic) are in blue, $\bN_{nm}$ with $\bN_{n'm'}$ (electric--electric) in red, the off-diagonal blocks coupling $\bM_{nm}$ with $\bN_{n'm'}$ (magnetic--electric) and $\bM_{n'm'}$ with $\bN_{nm}$ (electric--magnetic) are in orange and green, respectively. The elements coupling VSWFs of the same multipole order ($n = n'$) are of darker shade.}
\end{figure}
Outside a given scatterer, the total field $\bE_\mathrm{tot}(k\br) = \rbE_\mathrm{inc}(k\br) + \bE_\mathrm{sca}(k{\bf r})$ is partitioned into a known incident contribution $\rbE_\mathrm{inc}(k\br)$ and unknown scattered contribution $\bE_\mathrm{sca}(k\br)$. Both contributions are expanded in terms of VSWFs up to some multipole order $\nmax$,
\begin{align}
\rbE_\mathrm{inc}(k{\bf r}) & =  E \rbW(k\br) \rba, \\
\bE_\mathrm{sca}(k{\bf r}) & =  E \bW(k\br) \ba,
\end{align}
where $E$ corresponds to the incident field's amplitude (usually taken as unity, $E = |\rbE_\mathrm{inc} | = 1$), and $\rba\in \mathbb{C}^{\lmax}$, $\ba \in \mathbb{C}^{\lmax}$ are the incident and scattered coefficients, respectively. The association of $\rbE_\mathrm{inc}$ with regular (or \emph{incoming}) and $\bE_\mathrm{sca}$ with irregular (or \emph{outgoing}) VSWFs is a choice motivated by physical reasoning: (i) $\rbE_\mathrm{inc}(k\br)$ ought to be well defined everywhere within a finite distance from the origin, which rules out irregular VSWFs due to their singular behaviour at $\br = 0$; and (ii) $\bE_\mathrm{sca}(k\br)$ ought to satisfy the outgoing Sommerfeld radiation condition, requiring that $|\bE_\mathrm{sca}(\br)| \to 0$ as $1/r$ as $|\br| \to \infty$, which rules out regular VSWFs due to their divergence in the far field. Given the linearity of the governing Maxwell equations in linear media, the \tmatrix\ method expresses the linear dependence between $\rba$ and $\ba$, 
\begin{equation} \label{eqn:genTmatrix}
\ba = \bT\, \rba, \quad \mathrm{or} \quad a_{l} = \sum_{l'} T_{ll'}\widetilde{a}_{l'},
\end{equation}
where $\bT$ is the so-called "transition" or "transfer" \cite{MishchenkoTL02,StoutAL02} matrix (\tmatrix\ for short), which depends on the scatterer's characteristics at a given wavelength but is independent of illumination, encoded in $\rba$. For a spherically symmetric scatterer centred at the origin, $\bT$ is a diagonal matrix with the diagonal elements determined analytically by Mie theory. For other particle shapes, \TERMS\ requires that the \tmatrix\ be provided as input, with a format specified in App.~\ref{appendix:keywords} (keyword \keywd{TmatrixFiles}). Note that the \tmatrix\ may also represent the response of a composite scatterer comprising multiple particles; \TERMS\ can in fact calculate such a collective \tmatrix\ from individual one-body \tmatrices,\cite{StoutAL02, StoutAD08} and re-use it as input to simulate the scattering properties of a superstructure of such elements\cite{Fruhnert:2017aa}.
\subsection[Rotation and translation properties]{Transformation under rotation/translation of coordinates}
\begin{figure}
\center
\includegraphics[width=0.5\textwidth]{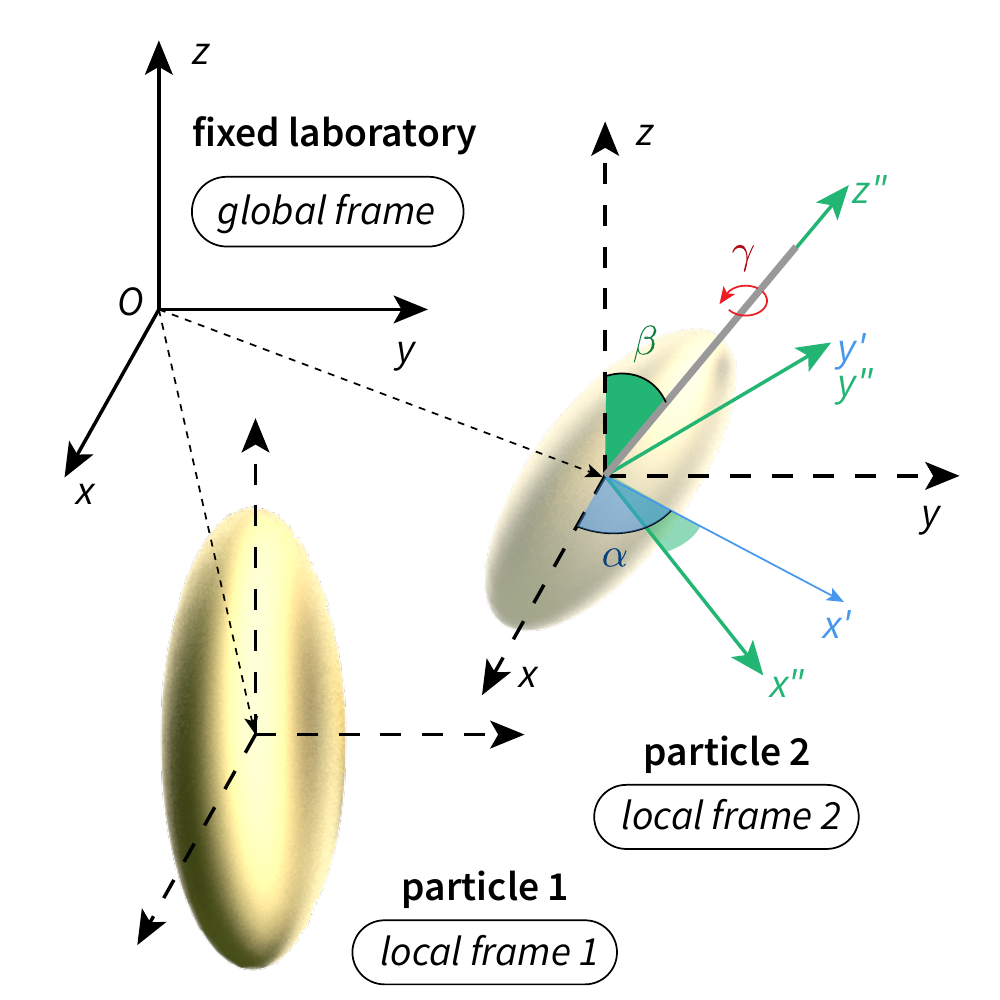}
\label{fig:frames}
\caption{Illustration of local and global reference frames for a cluster of particles and their associated \tmatrices.}
\end{figure}
The STM method requires transforming the series expansions of the fields from one origin to another, such as from the centre of one particle to a neighbour's, or to a common origin referred to as the \emph{global frame's}. Typically the \tmatrix\ of a nonspherical particle will have been calculated in a convenient orientation, e.g. for axisymmetric particles with symmetry axis along $z$, requiring rotations in changing reference frame as illustrated in Fig.~\ref{fig:frames}. In the following we summarise useful relations for the translation and rotation of VSWFs. We take the notational convention that expressions in local coordinate frames are specified with a superscript in brackets.
\subsubsection{Rotation}
Let $\br^{(1)} = (r,\theta^{(1)},\varphi^{(1)})$ and $\br^{(2)} = (r,\theta^{(2)},\varphi^{(2)})$ be the spherical polar coordinates of the same point $P$ in coordinate systems $1$ and $2$, respectively, sharing the same origin $O$. If coordinate system $2$ is obtained by rotating coordinate system $1$ through Euler angles $(\alpha,\beta,\gamma)$, here defined in the "$zyz$" convention\cite{MishchenkoTL02} with $0 \leq \alpha < 2\pi$, $0 \leq \beta \leq \pi$, and $0 \leq \gamma < 2\pi$, then
\begin{equation}\label{eqn:rotTrans}
\begin{split}
\psi_{nm}^{(2)}(k\br^{(2)})  = & \sum_{\mu=-n}^{n} \psi_{n\mu}^{(1)}(k\br^{(1)}) D_{\mu m}^{n}(\alpha,\beta,\gamma), \\
\psi_{nm}^{(1)}(k\br^{(1)})  = & \sum_{\mu=-n}^{n} \psi_{n\mu}^{(2)}(k\br^{(2)}) D_{\mu m}^{n}(-\gamma,-\beta,-\alpha),
\end{split}
\end{equation}
where $D_{\mu m}^{n} = e^{-i\mu\alpha}d_{\mu m}^{n}(\beta) e^{-im\gamma}$ and $d_{\mu m}^{n}$ are the Wigner $D$- and $d$-functions.\cite{MishchenkoTL02} Conveniently, $\widetilde{\psi}_{nm}$, $\rbM_{nm}$, $\rbN_{nm}$, $\bM_{nm}$ and $\bN_{nm}$ transform in exactly the same manner under rotation, so substituting $\psi_{nm}$ by a desired basis function in (\ref{eqn:rotTrans}) will give the appropriate expression (see equations (5.23)--(5.24) of Ref.~\citenum{MishchenkoTL02} for details). In our notation, $\bW^{(2)}(k\br^{(2)})$ in coordinate system $2$ is related to $\bW^{(1)}(k\br^{(1)})$ in coordinate system $1$ via
\begin{equation} \label{eqn:basisRot}
\bW^{(2)} = \bW^{(1)} \bR(\alpha,\beta,\gamma)
\end{equation}
where $\bR(\alpha,\beta,\gamma)$ is a unitary block-diagonal matrix (of size $\lmax \times \lmax$), satisfying 
\begin{equation}
\bR^{-1}(\alpha,\beta,\gamma) = \bR^{\dagger}(\alpha,\beta,\gamma) = \bR(-\gamma,-\beta,-\alpha)
\end{equation}
with matrix elements given by
\begin{equation}
R_{ll'}(\alpha,\beta,\gamma) = \delta_{ss'}\delta_{nn'} D_{m'm}^{n}(\alpha,\beta,\gamma), 
\end{equation}
where the index $l(s,n,m)$ is defined in (\ref{eqn:defLindex}). Note that (\ref{eqn:basisRot}) also applies to regular waves $\rbW$. Now, if a (regular or irregular) spherical wave expansion is described by a vector of coefficients $\ba^{(1)}$ in coordinate system $1$ and by $\ba^{(2)}$ in coordinate system $2$, then
\begin{equation} \label{eqn:rotMat}
\ba^{(2)} = \bR^{\dagger}(\alpha,\beta,\gamma) \ba^{(1)},
\end{equation}
which follows from equating the field expansions and using (\ref{eqn:basisRot}), i.e.
\begin{multline}
\bW^{(1)} \ba^{(1)} = \bW^{(2)} \ba^{(2)} = \bW^{(1)} \bR(\alpha,\beta,\gamma) \ba^{(2)} \\  \implies  \ba^{(1)} = \bR(\alpha,\beta,\gamma)\ba^{(2)}.
\end{multline}
Let us re-label coordinate system $1$ as $G$ to indicate a global, space-fixed reference frame, and coordinate system $2$ as $L$ for local frame, attached to a scatterer. A \tmatrix\ $\bT^{(L)}$ expressed in the local frame is transformed into $\bT^{(G)} = \bR \bT^{(L)} \bR^{\dagger}$ in the global frame, where $\bR(\alpha,\beta,\gamma)$ depends on the Euler angles $(\alpha,\beta,\gamma)$ that rotate frame $G$ onto frame $L$ (as opposed to $L$ onto $G$). To clarify, consider
\begin{align*}
\ba^{(L)} & =  \bT^{(L)}\rba^{(L)} \\
\bR^{\dagger} \ba^{(G)} & =  \bT^{(L)}\bR^{\dagger} \rba^{(G)} \\
\ba^{(G)} & =  \underbrace{\bR\bT^{(L)}\bR^{\dagger}}_{\bT^{(G)}} \rba^{(G)} \implies \bT^{(G)} = \bR \bT^{(L)} \bR^{\dagger}.
\end{align*}
If the scatterer is rotationally symmetric about the local $z$-axis, which is tilted by spherical polar angles $(\theta,\varphi)$ relative to the global $z$-axis, then $\alpha = \varphi$, $\beta = \theta$, and the value of $\gamma$ is irrelevant due to axial symmetry, so we can choose $\gamma = 0$ to have $\bT^{(G)}  = \bR(\varphi,\theta,0) \bT^{(L)} \bR(0,-\theta,-\varphi)$ (see Sec.\,5.2 of Ref.~\citenum{MishchenkoTL02} for details).
\subsubsection{Translation}
\begin{figure}
\center
\includegraphics[width=\columnwidth]{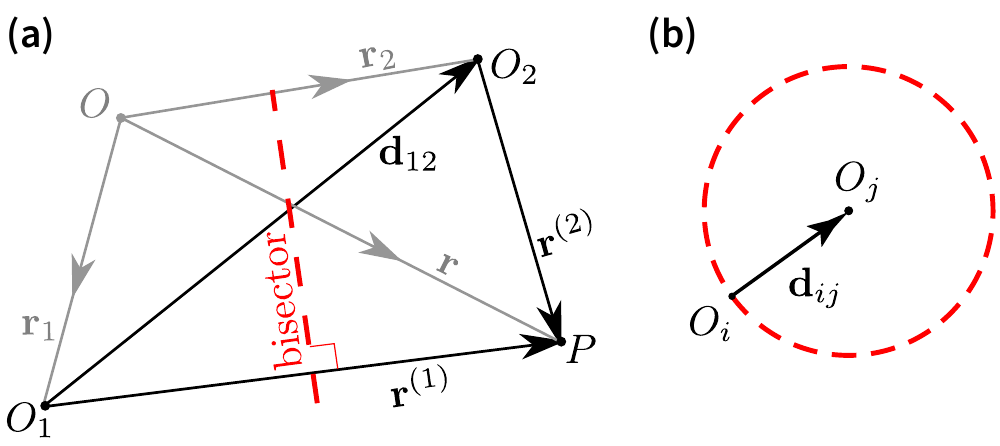}
\caption{\label{fig:transadd} (a) Illustration of how the coordinate vector of point $P$ is transformed from ${\bf r}^{(1)}$ to ${\bf r}^{(2)}$ when the origin is switched from $O_{1}$ to $O_{2}$. Dashed red line bisects the $O_{1}P$ edge of the $O_{1}PO_{2}$ triangle. (b) Illustration of how the singularity at $O_{i}$ exhibited by $\bW(k\br^{(i)})$ is spread over the surface of a ball with radius $d_{ij} = r_{ij}$ after translation to a target origin $O_{j}$ by displacement vector $\bd_{ij}$. The irregular basis remains irregular outside the ball, i.e.~$\bW(k\br^{(i)}) = \bW(k{\br^{(j)}})\rbO^{(j,i)}$ for $r^{(j)} > r_{ij}$, but is transformed into a \emph{regular} basis \emph{inside} the ball, i.e.~$\bW(k\br^{(i)}) = \rbW(k{\br^{(j)}})\bO^{(j,i)}$ for $r^{(j)} < r_{ij}$.}
\end{figure}

Consider a point $P$ with coordinates $\br^{(1)}$ in coordinate system $1$ with the origin at $O_{1}$. If we choose another origin $O_{2}$ displaced by $\bd_{12}$ from $O_{1}$, then the coordinates of $P$ relative to $O_{2}$ will be $\br^{(2)} = \br^{(1)} - \bd_{12}$, as illustrated in Fig.~\ref{fig:transadd}. The translation-addition theorem for vector spherical waves states that\cite{Chew:1990aa,Mackowski:2012ug}, in the limit $\nmax \to \infty$,
\begin{align}
\bW^{(1)}(k\br^{(1)}) & =  \begin{cases}
\bW^{(2)}(k\br^{(2)}) \rbO(k\bd_{12}), \text{ if } r^{(2)} > d_{12},  \\
\rbW^{(2)}(k\br^{(2)}) \bO(k\bd_{12}), \text{ if } r^{(2)} < d_{12},
\end{cases}  \label{eqn:iVTACS1} \\
\rbW^{(1)}(k\br^{(1)}) & =  \rbW^{(2)}(k\br^{(2)}) \rbO(k\bd_{12}), \label{eqn:rVTACS1}
\end{align}
where $\rbO(k\bd_{12})$ and $\bO(k\bd_{12})$ are ($\lmax \times \lmax$) matrices of regular and irregular translation-addition coefficients (TACs), respectively \cite{StoutAL02}. Note the conditional statement for irregular waves: the transformation depends on the relative length of $\br^{(2)}$ and $\bd_{12}$. In Fig.~\ref{fig:transadd}, an irregular basis centred at $O_{1}$ is mapped onto a regular basis centred at $O_{2}$ via the irregular TACs. However, if $O_{2}$ were to the left of the bisector, so that $r^{(2)} > d_{12}$, then the irregular basis centred at $O_{1}$ would be mapped onto an irregular basis centred at $O_{2}$ via the regular TACs. 

Note that $O_{1}$ and $O_{2}$ are themselves points with coordinates $\br_{1}$ and $\br_{2}$, respectively, in a common "global" frame with a fixed origin $O$. If we denote the global frame coordinates of $P$ by $\br$, then $\br^{(1)} = \br - \br_{1}$, $\br^{(2)} = \br - \br_{2}$, and $\bd_{12} = \br^{(1)} - \br^{(2)} = \br_{2} - \br_{1} =: \br_{21}$. Henceforth we follow Stout and coworkers\cite{StoutAL02,StoutAD08} and adopt the shorthand notation $\rbO^{(i,j)} := \rbO(k\br_{ij}) = \rbO(k\bd_{ji})$, and likewise for $\bO^{(i,j)}$, yielding
\begin{align}
\bW^{(i)} & =  \begin{cases}
\bW^{(j)} \rbO^{(j,i)},  \text{ if }  r^{(j)} > r_{ij},  \\
\rbW^{(j)} \bO^{(j,i)},  \text{ if }  r^{(j)} < r_{ij},
\end{cases}  \label{eqn:iVTACS2} \\
\rbW^{(i)} & =  \quad ~\rbW^{(j)} \rbO^{(j,i)}. \label{eqn:rVTACS2}
\end{align}
Note the reversal of indices in $\br_{ij} = \bd_{ji}$ and the minus sign in $\br_{ij} = -\bd_{ij}$; note that $d_{ij} = d_{ji} = r_{ij} = r_{ji} \geq 0$ in our notations.

To express the translation-addition theorem in terms of the coefficients (the $\ba$'s) of a VSWF expansion, multiply (from the right) both sides of equations (\ref{eqn:iVTACS2}) and (\ref{eqn:rVTACS2}) by column vector $\ba^{(i)}$, where the superscript ($i$) indicates where the VSWF expansion is centred. From inspection of the right-hand side we find that
\begin{align}
\ba^{(j)} & =  \rbO^{(j,i)} \ba^{(i)},  \text{ if } r^{(j)} > r_{ij}, \label{eqn:iVTACS3a}  \\
\rba^{(j)} & =  \bO^{(j,i)} \ba^{(i)},  \text{ if }  r^{(j)} < r_{ij}, \label{eqn:iVTACS3b} \\
\rba^{(j)} & =  \rbO^{(j,i)}\rba^{(i)}. \label{eqn:rVTACS3}
\end{align}
Note that (\ref{eqn:iVTACS3a}), (\ref{eqn:iVTACS3b}) and (\ref{eqn:rVTACS3}) are in a similar matrix-vector form to the rotation equation (\ref{eqn:rotMat}).
\subsubsection{Factorized translation (involving rotation)}
A general translation from centre $\br_{i}$ to another centre $\br_{j}$ by displacement vector $\bd_{ij} = (d_{ij},\theta_{ij},\varphi_{ij})$ can be separated into three steps:
\begin{enumerate}
\item Rotation of the local frame to align the $z$-axis with the $\bd_{ij}$ vector. In the $zyz$ convention, the appropriate Euler angles are $\alpha= \varphi_{ij}$, $\beta = \theta_{ij}$, and $\gamma = 0$.
\item Axial translation along the rotated local $z$-axis by $d_{ij}$.
\item Rotation of the local frame to realign the $z$-axis with the original orientation. The appropriate Euler angles are $\alpha' = -\gamma = 0$, $\beta' = -\beta = -\theta_{ij}$, and $\gamma' = -\alpha = - \varphi_{ij}$.
\end{enumerate}
This factorisation can be expressed in matrix form as 
\begin{equation}\label{eqn:rot-tranz-rot}
\bO^{(j,i)} = \bR(\varphi_{ij},\theta_{ij},0) \bO_{z}(d_{ij}) \bR(0,-\theta_{ij}, - \varphi_{ij})
\end{equation}
for the irregular case, where $\bO_{z}(d_{ij})$ represents the matrix of $z$-axial translation coefficients, many of which are zero due to the special case of axial translation along $z$. Note that the three aforementioned steps correspond to stepwise movement of the local axis, from the perspective of the initial point $i$, and the transformation corresponds to reading the matrix multiplication in (\ref{eqn:rot-tranz-rot}) from left to right; but the sequence of steps and the direction of movement is actually reversed from the perspective of the scatterer at the destination point. More importantly, since all three matrix-factors on the right-hand side of (\ref{eqn:rot-tranz-rot}) will contain many zeroes, operating on a vector of VSWF coefficients in a sequence of three steps can actually reduce the scaling of the net computational cost from $\sim \nmax^{4}$ to $\sim \nmax^{3}$, when the na\"ive matrix multiplication on each step is replaced by a custom operation that sums just over the relevant (non-zero) components. 

Another potential advantage of using (\ref{eqn:rot-tranz-rot}) is that, after obtaining the three factors for $\bO^{(j,i)}$, they can be recycled when computing the reverse translation $\bO^{(i,j)}$ to reduce the overall computational cost. First, beware that $\bO_{z}(-d_{ij})$ is \emph{not} the inverse of $\bO_{z}(d_{ij})$, and note that $\bO_{z}(d_{ij})$ is invariant to interchanging $i$ and $j$ ($d_{ij} = d_{ji} \geq 0$). Actually, $[\bO_{z}(d_{ij})]^{-1} = \bR(0,\pi,0) \bO_{z}(d_{ij}) \bR(0,-\pi,0)$, where $\bR(0,\pi,0)$ is block-diagonal and each block is \emph{anti}-diagonal. Second, since $\varphi_{ij} = \pi + \varphi_{ji}$ and $\theta_{ij} = \pi - \theta_{ji}$, $\bR(\varphi_{ji}, \theta_{ji},0)$ can be calculated from $\bR(\varphi_{ij}, \theta_{ij},0)$ using the symmetry relation $d_{\mu m}^{n}(\pi - \theta) = (-1)^{n-m}d_{-\mu m}^{n}(\theta)$ (see Eq.~B.7 in Ref.~\citenum{MishchenkoTL02}), so that $D_{\mu m}^{n}(\varphi_{ji},\theta_{ji},0) = (-1)^{n-m+1}D_{-\mu m}^{n}(\varphi_{ij},\theta_{ij},0)$, where the extra prefactor of $-1$ comes from multiplying by $e^{-i\pi} = -1$.    

Computing $\bO^{(i,j)}$ in general is more costly than the combined effort of computing $\bO_{z}$, $\bR$, \emph{and} $\bR^{-1}$. However, in our experience, performing matrix multiplication of the three factors in the sequence from the right ($\bR\bO_{z}\bR^{-1}$) does not seem to yield the expected result (matrix $\bO^{(i,j)}$), suggesting a numerical instability in this approach.
\subsection[STM for multiple scatterers]{Superposition \tmatrix\ for multiple scatterers}
\label{sec:STM}
\begin{figure}[!htpb]
\center
\includegraphics[width=\columnwidth]{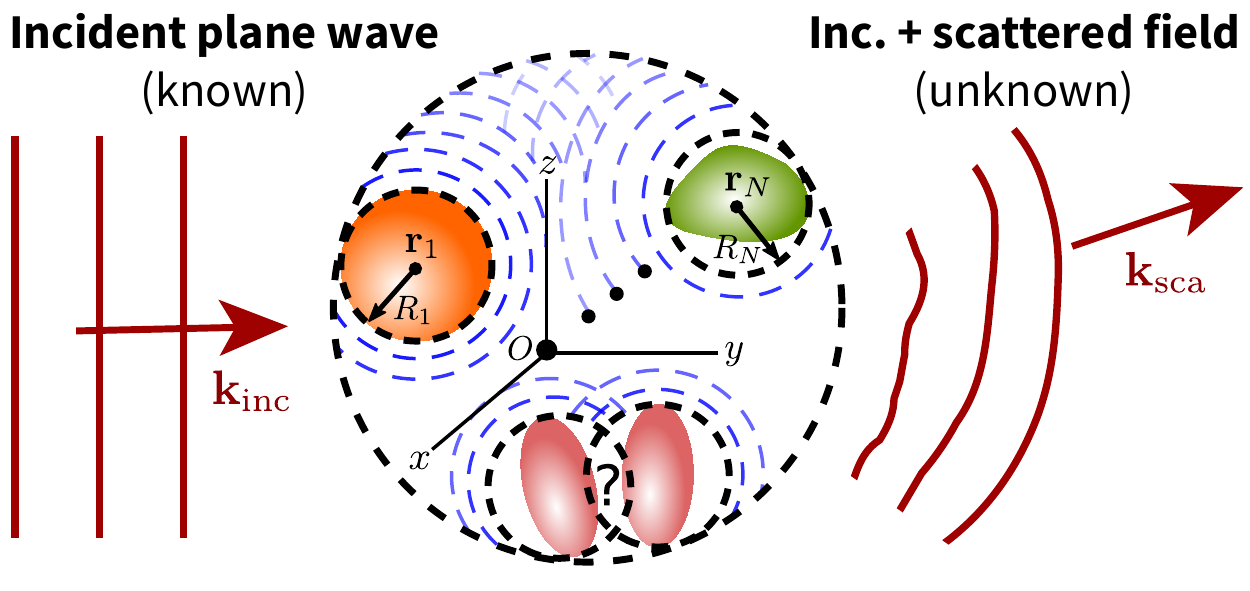}
\caption{Pictorial representation of light scatttering by a nanoparticle cluster. An incident plane wave with known wavevector ($\mathbf{k}_\mathrm{inc}$) and incident field ($\mathbf{E}_\mathrm{inc}$) is scattered by a cluster of $N$ particles centred at $\br_{1}, \br_{2}, \dots,\br_{N}$. Each particle scatters in response to the net incident field exciting it (partial waves illustrated in dashed blue). The self-consistent total field everywhere in space is the superposition of the incident field ($\mathbf{E}_\mathrm{inc}$), and of the collectively scattered field ($\mathbf{E}_\mathrm{sca}$). The scattered field is illustrated by distorted wavefronts outgoing from the cluster.}
\end{figure}
This section follows the treatments by Stout~\emph{et al.}~\cite{StoutAL02, StoutAD08} as well as similar discussions for multi-sphere clusters by Mackowski \& Mishchenko~\cite{Mackowski91, Mackowski94, Mackowski96,Mackowski:2012ug}.

For a cluster of $N$ scatterers (each of arbitrary shape), the collectively scattered field $\bE_\mathrm{sca}$ may be formally separated into additive contributions from the individuals, namely: 
\begin{align}
\bE_\mathrm{sca} (\br; k) &= \sum\limits_{j=1}^{N} \bE_{\mathrm{sca},j}(\br; k)\nonumber \\
{} &= E \sum\limits_{j=1}^{N} \bW^{(j)}(k{\bf r}^{(j)}) \bc_{j}^{(j)} \label{eqn:scaCoeffsJ}
\end{align}
where $\br^{(j)} = \br - \br_{j}$, $\br_{j}$ is the position of the $j$th scatterer in the global frame. Note that each partial field contribution $\bE_{\mathrm{sca},j}(\br)$ in (\ref{eqn:scaCoeffsJ}) is developed in terms of irregular waves centred at $\br_{j}$, as indicated in the superscript of $\br^{(j)}$ and the corresponding coefficients $\bc_{j}^{(j)}$. The centre of expansion need not necessarily be the centre of the particle associated with $\bE_{\mathrm{sca},j}(\br)$, so we still keep the subscript $j$ in $\bc_{j}^{(j)}$ as a label specifying the particle centre, which may seem redundant, but keep in mind that $\bc_{j}^{(i)}$ is well defined for $i \neq j$. It is also important to note that (\ref{eqn:scaCoeffsJ}) does not actually prescribe how exactly the collectively scattered field is partitioned among the individuals. The partitioning is to be determined self-consistently.
%
%
To set up a self-consistent system of linear equations, it is useful to define the excitation field $\rbE_{{\rm exc},j}(\br)$ for each scatterer $j$, and then develop it in terms of regular VSWFs centred at $\br_{j}$, i.e.
\begin{align}
\rbE_{{\rm exc},j}(\br) & := \rbE_\mathrm{inc}(\br) + \sum\limits_{\substack{l=1\\l \neq j}}^{N} \bE_{\mathrm{sca},l}(\br) \nonumber \\
 & =  E \rbW^{(j)}(k\br^{(j)}) \rba^{(j)} + E \sum\limits_{\substack{l=1\\l\neq j}}^{N} \bW^{(l)}(k\br^{(l)}) \bc_{l}^{(l)} \nonumber \\
 & =  E \rbW^{(j)}(k\br_{j}) \left( \rba^{(j)} + \sum\limits_{\substack{l=1\\l\neq j}}^{N} \underbrace{\bO^{(j,l)} \bc_{l}^{(l)}}_{\rbc_{j\leftarrow l}^{(j)}} \right)\label{eqn:excE}  \\
 {} & \text{ for } r^{(j)} < \min\|\br_{j} - \br_{l}\|\label{eqn:excEineq} \\
& =: E \rbW^{(j)} (k\br_{j}) \rbe_{j}^{(j)} \label{eqn:excCoeffsJ}
\end{align}
where $\rba^{(j)} = \rbO^{(j,0)} \rba$ contains the incident field coefficients and $\rbO^{(j,0)} := \rbO (k\br_{j})$. In the last equality of (\ref{eqn:excE}), $\bW^{(l)}$ is transformed into $\rbW^{(j)}$, with the corresponding coefficients given by $\rbc^{(j)}_{j\leftarrow l} = \bO^{(j,l)} \bc_{l}^{(l)}$, where the subscript $j\leftarrow l$ specifies the scattered field partition of scatterer $l$ developed (as a regular VSWF expansion) about particle $j$. Note the application of the translation-addition theorem clause that applies only inside the ball of radius $r_{jl}$ (for each $l\neq j$) centred at $\br_{j}$). This approach is strictly valid only if the translation ball fully contains the target scatterer's surface, where the boundary conditions are to be matched. While non-overlapping spherical scatterers are always guaranteed to satisfy this condition, elongated particles such as spheroids can be problematic, because the singularity sphere can cross the target scatterer's surface if it is sufficiently close (yet still not overlapping). These aspects are discussed in more details in Ref.~\citenum{schebarchov2019mind} (and references therein).

A self-consistent system of linear equations can now be obtained by requiring that
\begin{equation}
\bc_{j}^{(j)} = \bT_{j} \rbe_{j}^{(j)}, \label{eqn:Tmat1}
\end{equation}
where $\bT_{j}$ is the "one-body" \tmatrix\ characterising scatterer $j$, as defined above in Sec.~\ref{sec:tmatrix}. Note that (\ref{eqn:Tmat1}) reduces to (\ref{eqn:genTmatrix}) for a single scatterer ($j=N=1$) at the origin, because then $\rbe_{j}^{(j)} \mapsto \rba$, $\bc_{j}^{(j)} \mapsto \ba$, and $\bT_{j} \mapsto \bT$. From (\ref{eqn:excE}), (\ref{eqn:excCoeffsJ}), and (\ref{eqn:Tmat1}) we obtain an equation expressed just in terms of the field coefficients:
\begin{equation} \label{eqn:FunMultiScat}
\rbe_{j}^{(j)} =  \rba^{(j)} + \sum\limits_{\substack{i=1\\i\neq j}}^{N} \bO^{(j,i)} \bT_{i} \rbe_{i}^{(i)},
\end{equation}
which Stout~\emph{et al.} label as "the fundamental multiple scattering equation" (Eq.~9 in Ref.~\citenum{StoutAL02}). It is helpful to rewrite the linear system (\ref{eqn:FunMultiScat}) in block-matrix form:
\begin{widetext}
\begin{equation} \label{eqn:FunMultiScatMat}
\setlength{\tabcolsep}{-3pt}
\setlength{\arraycolsep}{2pt}
\renewcommand{\arraystretch}{0.5}
\small
\begin{bmatrix}
\bI &  -\bO^{(1,2)} \bT_{2} & \cdots & -\bO^{(1,N)}\bT_{N} \\
-\bO^{(2,1)} \bT_{1} & \bI & \cdots & -\bO^{(2,N)} \bT_{N} \\ 
\vdots & \vdots & \ddots & \vdots \\
-\bO^{(N,1)} \bT_{1} & -\bO^{(N,2)} \bT_{2} & \cdots & \bI
\end{bmatrix}
\begin{pmatrix}
\rbe_{1}^{(1)} \\
\rbe_{2}^{(2)} \\
\vdots \\
\rbe_{N}^{(N)}
\end{pmatrix}
=
\begin{pmatrix}
\rba^{(1)} \\
\rba^{(2)} \\
\vdots \\
\rba^{(N)} \\
\end{pmatrix},
\end{equation}%
\end{widetext}
which is in the standard form $\mathbf{Ax} = \mathbf{b}$, with $\mathbf{x}$ the unknown, and $\mathbf{b}$ a known input source. The solution $\bx$ gives all the $\rbe_{i}^{(i)}$'s for a given $\rba$ and $\bT_{i}$ ($i=1,\dots,N$), from which we can determine all the $\rbc_{i}^{(i)}$'s using (\ref{eqn:Tmat1}). Alternatively, we can use (\ref{eqn:Tmat1}) to substitute the excitation field coefficients for the scattered field coefficients and, assuming the one-body \tmatrices\ are invertible, obtain

\begin{widetext}
\begin{equation} \label{eqn:FunMultiScat2}
\bT_{j}^{-1}\bc_{j}^{(j)} - \sum\limits_{\substack{i=1\\i\neq j}}^{N} \bO^{(j,i)}  \bc_{i}^{(i)}  = \rba^{(j)},
\end{equation}
\begin{equation} \label{eqn:FunMultiScatMat2}
\begin{bmatrix}
\bT_{1}^{-1} &  -\bO^{(1,2)} & \cdots & -\bO^{(1,N)} \\
-\bO^{(2,1)} & \bT_{2}^{-1} & \cdots & -\bO^{(2,N)} \\ 
\vdots & \vdots & \ddots & \vdots \\
-\bO^{(N,1)} & -\bO^{(N,2)} & \cdots & \bT_{N}^{-1}
\end{bmatrix}
\begin{pmatrix}
\bc_{1}^{(1)} \\
\bc_{2}^{(2)} \\
\vdots \\
\bc_{N}^{(N)}
\end{pmatrix}
=
\begin{pmatrix}
\rba^{(1)} \\
\rba^{(2)} \\
\vdots \\
\rba^{(N)} \\
\end{pmatrix}.
\end{equation}%
%
To avoid involving the matrix inverses $\bT_{j}^{-1}$, we can also rearrange the linear system into

\begin{equation} \label{eqn:FunMultiScat3}
\bc_{j}^{(j)} -\bT_{j} \sum\limits_{\substack{i=1\\i\neq j}}^{N} \bO^{(j,i)}  \bc_{i}^{(i)}  = \bT_{j} \rba^{(j)},
\end{equation}
\begin{equation} \label{eqn:FunMultiScatMat3}
\begin{bmatrix}
\bI &  -\bT_{1} \bO^{(1,2)} & \cdots & -\bT_{1}\bO^{(1,N)} \\
-\bT_{2} \bO^{(2,1)} & \bI & \cdots & -\bT_{2}\bO^{(2,N)}  \\ 
\vdots & \vdots & \ddots & \vdots \\
- \bT_{N} \bO^{(N,1)}  & -\bT_{N} \bO^{(N,2)}  & \cdots & \bI
\end{bmatrix}
\begin{pmatrix}
\bc^{(1)}_{1} \\
\bc^{(2)}_{2} \\
\vdots \\
\bc^{(N)}_{N}
\end{pmatrix}
=
\begin{pmatrix}
\ba_{1}^{(1)} \\
\ba_{2}^{(2)} \\
\vdots \\
\ba_{N}^{(N)} \\
\end{pmatrix},
\end{equation}
\end{widetext}

where $\ba_{j}^{(j)} = \bT_{j} \rba^{(j)}$ corresponds to irregular series coefficients for the scattered field of particle $j$ in isolation (from all the other $N-1$ particles).\footnote{Note that (\ref{eqn:FunMultiScat3}) is equivalent to Mackowski's Eq.~4 of Ref.~\citenum{MackowskiM11}, though in Mackowski \& Mishchenko's earlier papers the same equation (Eq.~13 of Ref.~\citenum{Mackowski94} and Eq.~3 of Ref.~\citenum{Mackowski96}) has a plus sign instead of the minus, which may be entirely due to a minus sign featuring in the incident field expansion (see Eq.~4 in Ref.~\citenum{Mackowski94}). This minus sign is absent in Eq.~2 of the more recent Ref.~\citenum{MackowskiM11}, and the subsequent Eq.~4 matches our equation (\ref{eqn:FunMultiScat3}).}

Note that (\ref{eqn:FunMultiScatMat}), (\ref{eqn:FunMultiScatMat2}), and (\ref{eqn:FunMultiScatMat3}) are all in the form of a general matrix equation $\mathbf{A}\bx = \mathbf{b}$, which can be solved for the column vector(s) $\bx$ without inverting the matrix $\mathbf{A}$. However, formal inversion is necessary when seeking \emph{collective} \tmatrix\ constructions, which describe the entire cluster's response to arbitrary incident fields, and can notably provide analytical formulas for orientation-averaged quantities. A number of collective \tmatrices\ are defined in the next section, following the different treatments of Stout and Mackowski \& Mishchenko.
%
\subsection{Collective \tmatrix\ constructions}
The system of coupled matrix equations in (\ref{eqn:FunMultiScatMat2}) can be solved for $\bc_{j}^{(j)}$ by inverting the matrix to obtain
\begin{align}\label{eqn:scatCentTMat}
\bc^{(j)}_{j} = \sum_{i=1}^{N} \bT^{(j,i)} \rba^{(i)}
 {} & = \underbrace{\left(\sum_{i=1}^{N} \bT^{(j,i)} \rbO^{(i,0)}\right)}_{:=\mathrm{Mackowski's}~ \bT^{(j)}_{M}} \rba \\
 {} & = \underbrace{\left(\sum_{i=1}^{N} \bT^{(j,i)} \rbO^{(i,j)}\right)}_{:=\mathrm{Stout's}~ \bT_{S}^{(j)}} \rba^{(j)},
\end{align}
where $\bT^{(j,i)}$ represent what we may call "pairwise \tmatrices", expressing the portion of the scattered field from particle $j$ in response to its excitation by particle $i$; the $\bT^{(j,i)}$ matrices are arranged and defined as follows

\begin{widetext}
\begin{align}
\left[\bT^{(j,i)}\right] & = 
\begin{bmatrix}
\bT^{(1,1)} & \bT^{(1,2)} & \cdots & \bT^{(1,N)} \\
\bT^{(2,1)} & \bT^{(2,2)} &
\cdots & \bT^{(2,N)} \\
\vdots & \vdots & \ddots & \vdots \\
\bT^{(N,1)} & \bT^{(N,2)} & \cdots & \bT^{(N,N)}
\end{bmatrix}
 =
\begin{bmatrix}
\bT_{1}^{-1} &  -\bO^{(1,2)} & \cdots & -\bO^{(1,N)} \\
-\bO^{(2,1)} & \bT_{2}^{-1} & \cdots & -\bO^{(2,N)} \\ 
\vdots & \vdots & \ddots & \vdots \\
-\bO^{(N,1)} & -\bO^{(N,2)} & \cdots & \bT_{N}^{-1}
\end{bmatrix}^{-1} \label{eqn:TjiMatrixDefinition} \\
& =  
\begin{bmatrix}
\bT_{1} &  0 & \cdots & 0 \\
0 & \bT_{2} & \cdots & 0 \\ 
\vdots & \vdots & \ddots & \vdots \\
0 & 0 & \cdots & \bT_{N}
\end{bmatrix}
\begin{bmatrix}
\bI &  -\bO^{(1,2)} \bT_{2} & \cdots & -\bO^{(1,N)} \bT_{N} \\
-\bO^{(2,1)}\bT_{1} & \bI & \cdots & -\bO^{(2,N)} \bT_{N} \\ 
\vdots & \vdots & \ddots & \vdots \\
-\bO^{(N,1)}\bT_{1} & -\bO^{(N,2)} \bT_{2} & \cdots & \bI
\end{bmatrix}^{-1} \label{eqn:TjiMatrixFactored1}\\
& = 
\begin{bmatrix}
\bI &  - \bT_{1} \bO^{(1,2)} & \cdots & -\bT_{1} \bO^{(1,N)} \\
-\bT_{2} \bO^{(2,1)} & \bI & \cdots & -\bT_{2} \bO^{(2,N)} \\ 
\vdots & \vdots & \ddots & \vdots \\
-\bT_{N} \bO^{(N,1)} & -\bT_{N} \bO^{(N,2)} & \cdots & \bI
\end{bmatrix}^{-1}
\begin{bmatrix}
\bT_{1} &  0 & \cdots & 0 \\
0 & \bT_{2} & \cdots & 0 \\ 
\vdots & \vdots & \ddots & \vdots \\
0 & 0 & \cdots & \bT_{N} \label{eqn:TjiMatrixFactored2}
\end{bmatrix}
\end{align}
\end{widetext}
with the last two lines merely showing how the one-body \tmatrices\ can be factored out in two different ways (left or right).

The $\bT^{(j,i)}$ matrices provide a complete and exact solution to the multiple scattering problem. Crucially, they retain all the information required to calculate fields at any point within or outside the cluster (except within the Rayleigh Hypothesis region for nonspherical scatterers). Stout~\emph{et al.}~denotes these matrices "scatterer-centred transfer matrices" (see Eq.~12 in Ref.~\citenum{StoutAL02}), while Mackowski \& Mishchenko refer to them as "sphere-centred" (see Eq.~16 in Ref.~\citenum{Mackowski94} and Eq.~4 in Ref.~\citenum{Mackowski96}). Arguably, both appellations are equally applicable to $\bT^{(j)}$, which Mackowski \& Mishchenko define one way (see Eq.~61 in Ref.~\citenum{Mackowski96}) without giving a particular name, while Stout~\emph{et al.} define $\bT^{(j)}$ differently and call it "individual $N$-body transfer matrices" (see equations.~14 and 27 in Ref.~\citenum{StoutAL02}). The difference between both is explicitly stated in Eq.~(\ref{eqn:scatCentTMat}), with Stout's $\bT_{S}^{(j)}$ retaining expansions from each scatterer's origin, and Mackowski \& Mishchenko's $\bT_{M}^{(j)}$ collapsing all expansions to a common origin $O$. Note that neither definition should be confused with the one-body \tmatrices\ $\bT_{j}$ of equation (\ref{eqn:Tmat1}).

Mackowski \& Mishchenko additionally consider the collective scattering coefficients $\ba$ for the irregular VSWF expansion about the common origin of the whole cluster, i.e.
\begin{align}
 \bE_\mathrm{sca} (\br; k) = & E \bW(k\br) \ba\label{eqn:fcl}\\
  {} = & E \bW(k\br) \left(\sum\limits_{j=1}^{N} \rbO^{(0,j)}\bc^{(j)}\right),\nonumber\\ {}& \text{for}  \|\br\| > \max\|\br_{j}\|\nonumber,
\end{align}
where the second equality relies on a particular clause of the translation-addition theorem, which is valid only outside of the smallest circumscribed sphere (encompassing all $N$ scatterers) centred at the global frame's origin. From (\ref{eqn:fcl}) and (\ref{eqn:scatCentTMat}) we have
\begin{align} 
\ba &= \sum_{j=1}^{N} \rbO^{(0,j)} \bc^{(j)}_{j}\nonumber \\
&= \sum_{j=1}^{N} \rbO^{(0,j)} \sum_{i=1}^{N} \bT^{(j,i)} \rba^{(i)}\nonumber \\
&= \sum_{j=1}^{N} \rbO^{(0,j)} \bT_{M}^{(j)} \rba,\label{eqn:contraction}
\end{align}
where $\ba$ and $\rba$ are now related as in (\ref{eqn:genTmatrix}), providing expressions for the collective \tmatrix\ of the entire cluster
\begin{align} 
\bT &= \sum_{j=1}^{N}\sum_{i=1}^{N} \rbO^{(0,j)} \bT^{(j,i)} \rbO^{(i,0)}\nonumber \\
&= \sum_{j=1}^{N} \rbO^{(0,j)} \bT_{M}^{(j)} \nonumber \\
 & = \sum_{j=1}^{N} \rbO^{(0,j)} \bT_{S}^{(j)} \rbO^{(j,0)},\label{eqn:Tcl}
\end{align}
where in the last equality we used the fact that $\bT^{(j)}_{M} = \bT_{S}^{(j)} \rbO^{(j,0)}$. Mackowski \& Mishchenko refer to the collective $\bT$ in (\ref{eqn:Tcl}) as the "cluster-centred" \tmatrix\ (see Eq.~19 in Ref.~\citenum{Mackowski94}, Eq.~64 in Ref.~\citenum{Mackowski96}, and Eq.~29 in Ref.~\citenum{MackowskiM11}). Note that (\ref{eqn:contraction}) is valid only outside of the cluster's smallest circumscribed sphere centred at the common origin; this collective \tmatrix\ does not allow the calculation of near-fields between particles.\cite{Mackowski96, MackowskiM11}

In the \TERMS\ program, when $\keywd{Scheme} \neq 0$ the collective \tmatrix\ is calculated in the subroutine \keywd{contractTmat} of the \module{multiscat} module; it is used to calculate orientation-averaged far-field cross-sections. However, if the keyword \keywd{ScattererCentredCrossSections} is included in the input file, the collective $\bT$ will not be calculated and the program will be using alternative orientation-averaging formulas based on particle-centred \tmatrices\ $\bT^{(i,j)}$ instead.

\subsection{Far-field cross-sections}
In $\keywd{Mode}=2$ the program calculates far-field cross-sections for the given incident field direction(s) and four polarisations (two linear, two circular), as well as their average over the full solid angle using analytical formulas.

\subsubsection{Fixed orientation cross-sections}
After solving for the particle-centred coefficients $\bc_{j}^{(j)}$ for a given $\rba$ and $\bT_{j}$'s (where $j=1,\dots, N$), the corresponding fixed-orientation extinction ($\sigma_{\rm ext}$), scattering ($\sigma_\mathrm{sca}$), and absorption ($\sigma_{\rm abs}$) cross-sections can be calculated. Here we state formulae for $\sigma_{\rm ext}$ and $\sigma_\mathrm{sca}$ expressed in terms of origin- and particle-centred coefficients, without derivation but with references to the previously-cited literature. For a non-absorbing incident medium (real-valued wavenumber $\km$, as must be the case throughout \TERMS), $\sigma_{\rm abs}$ can be inferred using energy conservation: $\sigma_{\rm abs} = \sigma_{\rm ext} - \sigma_\mathrm{sca}$.

Fixed-orientation extinction cross-sections are calculated using
\begin{align} 
\sigma_{\rm ext}  = & -\frac{\Re\{\rba ^{\dagger} \ba \}}{\km^{2}}  = -\frac{1}{\km^{2}}  \sum_{l=1}^{\lmax} \Re \left \{ \widetilde{a}_{l}^{*} a_{l} \right\} \label{eqn:csExt0} \\
{} = & -\frac{1}{\km^{2}} \Re \left\{ \sum_{j=1}^{N} \rba^{(j)\dagger}\bc^{(j)}\right\} \\
{} = & -\frac{1}{\km^{2}} \sum_{j=1}^{N}\sum_{l=1}^{\lmax} \Re \left\{ \widetilde{a}^{(j)*}_{l} c^{(j)}_{l} \right \} \label{eqn:csExt1}\\
=: & \sum_{j=1}^{N} \sigma_{\rm ext}^{(j)},
\end{align}
where $\Re\{ \dots \}$ indicates taking the real part of the quantity inside the braces, $\rba^{(j)\dagger}$ is conjugate transpose of the column vector $\rba^{(j)}$, $\widetilde{a}^{(j)*}_{l}$ is the complex conjugate of the vector component $a^{(j)}_{l}$, and $\km$ is the wavenumber in the incident medium. Note that, since $\Re\{ab^{*}\} = \Re\{ a^{*}b\}$ for any complex numbers $a$ and $b$, (\ref{eqn:csExt0}) is equivalent to Eq.~H.65 of Ref.~\citenum{LeRuE08} and Eq.~5.18a of Ref.~\citenum{MishchenkoTL02} (provided $|\bE_{0}^\mathrm{inc}|^2 = E = 1$). Equation (\ref{eqn:csExt1}) is taken from Eq.~29 of Ref.~\citenum{StoutAD08}, which follows from simplifying Eq.~43 of Ref.~\citenum{StoutAL02} and substituting into Eq.~42 of the same reference. Also note that $\sigma_{\rm ext}$ is (formally) separable into additive contributions from individual scatterers, i.e. $\sigma_{\rm ext} = \sum_{j}\sigma_{\rm ext}^{(j)}$.

Fixed-orientation scattering cross-sections can be calculated using
\begin{align}
\sigma_\mathrm{sca}  = &  \frac{|\ba |^{2}}{\km^2} = \frac{1}{\km^{2}}  \sum_{l=1}^{\lmax} a_{l}^{*} a_{l} \label{eqn:sigSca0} \\
 = & \frac{1}{\km^2} \sum_{j=1}^{N} \sum_{i=1}^{N} \ba^{(j)\dagger} \rbO^{(j,i)} \ba^{(i)}\\
 = & \frac{1}{\km^2} \sum_{j=1}^{N} \sum_{i=1}^{N} \sum_{l=1}^{\lmax} \sum_{l'=1}^{\lmax} a_{l}^{(j)*} \widetilde{O}_{ll'}^{(j,i)} a_{l'}^{(i)}\label{eqn:sigSca1}\\
=: & \sum_{j=1}^{N} \sigma_\mathrm{sca}^{(j)}, 
\end{align}
where $|\ba |^{2} = \ba^{\dagger}\ba$. Note that (\ref{eqn:sigSca0}) is equivalent to Eq.~H.64 of Ref.~\citenum{LeRuE08} and Eq.~5.18b of Ref.~\citenum{MishchenkoTL02} (provided $|\bE_{0}^\mathrm{inc}|^2 = E = 1$); while (\ref{eqn:sigSca1}) is from Eq.~29 of Ref.~\citenum{StoutAD08}, which follows from substituting Eq.~45 of Ref.~\citenum{StoutAL02} into Eq.~42 of the same reference. 

In \TERMS, with $\keywd{Mode}=2$ all the far-field cross-sections are calculated in the main subroutine \subroutine{spectrumFF}. If the keyword \keywd{ScattererCentredCrossSections} is included in the input file, fixed orientation cross-sections, are calculated using particle-centred coefficients via the subroutine \subroutine{calcCsStout} (equations \ref{eqn:csExt1} and \ref{eqn:sigSca1} are implemented in this subroutine). Otherwise, they are calculated using origin-centred coefficients via the subroutine \subroutine{calcCs}.
\subsubsection{Orientation averaged cross-sections}
One attractive feature of the \tmatrix\ method is that it provides relatively simple means of calculating orientation-averaged cross-sections, herein denoted by $\langle \sigma_{\rm ext}\rangle$, $\langle \sigma_\mathrm{sca}\rangle$ and $\langle \sigma_{\rm abs}\rangle$; these are often used to describe a randomly oriented scatterer, or, equivalently, light incident from a random direction\cite{Mishchenko:2017uf}. Note that in \TERMS\ the orientation averaging applies to the cluster as a whole, not to individual particles within the cluster: they are considered rigidly held together (attached on a template, in practice). As with the fixed-orientation cross-sections, orientation averages can be calculated either from the origin-centred collective \tmatrix\ $\bT$, or from the particle-centred \tmatrices\ $\bT^{(i,j)}$.
\begin{align}
\langle \sigma_{\rm ext}\rangle  = & -\frac{2\pi}{\km^{2}} \Re \left\{ \tr (\bT)\right\} \label{Eq:OAext_colT}\\
 = &  -\frac{2\pi}{\km^{2}} \sum_{j} \sum_{k} \Re \left\{ \tr\left(\bT^{(j,k)}\rbO^{(k,j)} \right) \right\}  \label{Eq:OAext_Tij}\\
 =: & \sum_{j} \langle \sigma_{{\rm ext},j} \rangle 
 \label{eq:cextoa}
\end{align}
\begin{align}
\langle \sigma_\mathrm{sca}\rangle  & =  \frac{2\pi}{\km^{2}}  \tr \left(\bT^{\dagger}\bT \right) \label{Eq:OAsca_colT} \\
 {} & =  \frac{2\pi}{\km^{2}} \sum_{j} \sum_{k} \tr \Bigg( \left[\sum_{l}\rbO^{(k,l)}\bT^{(l,j)}\right]^{\dagger} \label{Eq:OAsca_Tij} \\
 {} &   \hskip9em\!\left[\sum_{i}\bT^{(k,i)}\rbO^{(i,j)}\right] \Bigg)\nonumber\\
 {}  & =:  \sum_{j} \langle \sigma_{\mathrm{sca},j} \rangle 
  \label{eq:cscaoa}
\end{align}
The cluster's absorption cross-section can be calculated from energy conservation, 
\begin{equation} 
\langle \sigma_\mathrm{abs}\rangle = \langle \sigma_\mathrm{ext}\rangle- \langle \sigma_\mathrm{sca}\rangle .\label{Eq:OAabs_Tcol}
\end{equation} 
Alternatively, it may also be calculated directly from the flux of the Poynting vector of the total field entering the surface of each individual particle. This provides the physically-meaningful portion of energy absorbed within each scatterer $j$, and their sum adds up to the total absorption cross-section for the cluster. Following Stout\cite{StoutAL02} and restricting ourselves to non-magnetic, homogeneous spheres,
\begin{align}
\langle \sigma_\mathrm{abs}\rangle  & = \frac{2\pi}{\km^{2}} \sum_{j} \sum_{k} \sum_{l} \tr \Bigg( \left[\bT^{(j,k)}\right]^{\dagger} \Gamma^j \label{Eq:OAabs_Tij}\\
 {} &   \hskip9em\! \bT^{(j,l)}\rbO^{(l,k)} \Bigg)\nonumber\\
 {}  & =: \sum_{j} \langle \sigma_{{\rm abs},j} \rangle 
\end{align}   
where the absorption matrix $\Gamma^j$ is of the form 
\begin{equation}
\Gamma^j = 
\begin{bmatrix}
C^j & 0 \\
0   & D^j
\end{bmatrix}.
\end{equation}
$C^j$ and $D^j$ are diagonal matrices with matrix elements
\begin{align}
C_n^j = \frac{\Re\left[i\rho_j\psi_n^*(\rho_j\chi_j)\psi_n^{'}(\rho_j\chi_j) \right]}{\vert\psi_n(\rho_j\chi_j)\psi_n^{'}(\chi_j)-\rho_j\psi_n^{'}(\rho_j\chi_j)\psi_n(\chi_j) \vert^2} \\
D_n^j = \frac{\Re\left[i\rho_j^*\psi_n^*(\rho_j\chi_j)\psi_n^{'}(\rho_j\chi_j) \right]}{\vert\rho_j\psi_n(\rho_j\chi_j)\psi_n^{'}(\chi_j)-\psi_n^{'}(\rho_j\chi_j)\psi_n(\chi_j) \vert^2}  
\end{align}
where $\psi_n(x)$ are Ricatti-Bessel functions: $\psi_n(x) = x j_n(x)$, $\chi_j = kR_j$, and $\rho_j = k_j / \km$. $\km$ is the wavenumber in the incident medium, $R_j$ the radius of sphere $j$, and $k_j$ the wavenumber inside (homogeneous) sphere $j$.
 
When $\keywd{Scheme} = 1\text{ or } 2$ and the input file requests \keywd{ScattererCentredCrossSections}, the orientation-averaged cross-sections are calculated in the subroutine \subroutine{calcOaStout} which uses particle-centred \tmatrices\ $\bT^{(i,j)}$ (Eqs.~\ref{Eq:OAext_Tij}, \ref{Eq:OAsca_Tij}, \ref{Eq:OAabs_Tij}). Per-particle orientation-averaged absorption is currently only returned for homogeneous spheres, as the generalisation of Eq.~\ref{Eq:OAabs_Tij} to arbitrary scatterers is not yet available.

In other cases, orientation-averaged cross-sections are calculated in the subroutine \subroutine{calcOAprops}, which uses the common-origin collective \tmatrix\ $\bT$ (Eqs.~\ref{Eq:OAext_colT}, \ref{Eq:OAsca_colT}, \ref{Eq:OAabs_Tcol}). Note that these calculations based on collective $\bT$ are much faster than those based on particle-centred \tmatrices\ $\bT^{(i,j)}$.
\subsubsection{Circular dichroism} 
\label{sec:CD}
Circular dichroism (CD) is defined as the difference between the optical properties of the structure under left and right circularly polarised incident fields. Its calculation is more natural when VSWFs are expressed in the "helicity" basis, related to the standard "parity" (\textsc{te, tm}) basis via the helicity operator ($\Lambda= \frac{\nabla \times}{k}$) leading to\citep{Suryadharma18},
\begin{align}
		& \bhelB_{R,nm}=\frac{1}{\sqrt{2}}(\bM_{nm}-\bN_{nm}),\quad \Lambda\bhelB_{R,nm} =-\bhelB_{R,nm} \label{eq:R-helicity basis} \\
		& \bhelB_{L,nm}=\frac{1}{\sqrt{2}}(\bM_{nm}+\bN_{nm}), \quad \Lambda\bhelB_{L,nm}=\bhelB_{L,nm}, \label{eq:L-helicity basis}
	\end{align}
where the subscripts ($R$) and ($L$) refer to right and left circularly polarised light. The corresponding \tmatrix\ describes the scattering of circularly-polarised incident fields in the helicity basis.

Using these definitions the relation between the \tmatrix\ blocks in parity and helicity bases reads
\begin{align}
	\begin{bmatrix}
		T_{\tinyLL} & T_{\tinyLR} \\
		T_{\tinyRL} & T_{\tinyRR}
	\end{bmatrix}
	=\frac{1}{2}\begin{bmatrix}
		I & I \\
		I & -I
	\end{bmatrix}
	\begin{bmatrix}
		T_{11} & T_{12}\\
		T_{21} & T_{22}
	\end{bmatrix}
	\begin{bmatrix}
		I & I \\
		I & -I
	\end{bmatrix}, \label{eq:Tcp2}
\end{align}
where $I$ is the identity matrix with the same size as the 4 matrix blocks ($T_{11}$, etc.). The orientation averaged cross-sections for a specific (L) or (R) polarisation can be obtained from Eqs.~(\ref{eq:cextoa}), (\ref{eq:cscaoa}), by restricting the coefficients of the incident field to one helicity,\cite{Suryadharma18}
\begin{align}
\langle \sigma_\mathrm{sca}\rangle_L  & =  \frac{4\pi}{\km^{2}}  \tr \left(\bT_{LL}^{\dagger}\bT_{LL}+\bT_{RL}^{\dagger}\bT_{RL} \right)\\
\langle \sigma_\mathrm{ext}\rangle_L  & =  \frac{4\pi}{\km^{2}} \Re \left[\tr \left(\bT_{LL}\right)\right]\\
\langle \sigma_\mathrm{abs}\rangle_L  & = \frac{4\pi}{\km^{2}}  \Re  \left[\tr \left(\bT_{LL}(\textbf{I}-\bT_{LL}^{\dagger})-\bT_{RL}^{\dagger}\bT_{RL} \right)\right] 
\end{align}
with simple changes $L\leftrightarrow R$ for R polarisation. Circular dichroism is then obtained as the difference between (L) and (R) cross-sections.

The subroutine \subroutine{calcOAprops} implements these formulas, calculated for $\keywd{Scheme} \neq 0$.

\subsubsection{Stokes scattering vector and phase matrix}
%
Some light scattering applications require characterising the angular and polarisation characteristics of the scattered field for a specified incident plane wave. \terms\ uses the Stokes vector formalism to describe such situations in $\keywd{Mode}=3$, following Ref.~\citenum{MishchenkoTL02}. From the incident electric field $\mathbf{E}_{0}$, the components of the incident Stokes vector read
\begin{equation}
\bI=
\begin{bmatrix} I\\ Q\\ U\\ V\\ \end{bmatrix}
=\dfrac{1}{2}\sqrt{\dfrac{\varepsilon}{\mu}}
\begin{bmatrix}
E_{0\theta}E_{0\theta}^*+E_{0\varphi}E_{0\varphi}^* \\
E_{0\theta}E_{0\theta}^*-E_{0\varphi}E_{0\varphi}^* \\
-2\Re(E_{0\theta}E_{0\varphi}^*)\\
2\Im(E_{0\theta}E_{0\varphi}^*)\\
\end{bmatrix}
\end{equation}
The $4\times 4$ phase matrix $\bZ$ relates incident and scattered field Stokes vectors, with the following expressions,
\begin{align}
Z_{11}&=\tfrac{1}{2}(|S_{11}|^2+|S_{12}|^2+|S_{21}|^2+|S_{22}|^2)\\
Z_{12}&=\tfrac{1}{2}(|S_{11}|^2-|S_{12}|^2+|S_{21}|^2-|S_{22}|^2)\\
Z_{13}&=-\Re(S_{11}S_{12}^*+S_{22}S_{21}^*)\\
Z_{14}&=-\Im(S_{11}S_{12}^*-S_{22}S_{21}^*)\\
Z_{21}&=\tfrac{1}{2}(|S_{11}|^2+|S_{12}|^2-|S_{21}|^2-|S_{22}|^2)\\
Z_{22}&=\tfrac{1}{2}(|S_{11}|^2-|S_{12}|^2-|S_{21}|^2+|S_{22}|^2)\\
Z_{23}&=-\Re(S_{11}S_{12}^*-S_{22}S_{21}^*)\\
Z_{24}&=-\Im(S_{11}S_{12}^*+S_{22}S_{21}^*)\\
Z_{31}&=-\Re(S_{11}S_{21}^*+S_{22}S_{12}^*)\\
Z_{32}&=-\Re(S_{11}S_{21}^*-S_{22}S_{12}^*)\\
Z_{33}&=\Re(S_{11}S_{22}^*+S_{12}S_{21}^*)\\
Z_{34}&=\Im(S_{11}S_{22}^*+S_{21}S_{12}^*)\\
Z_{41}&=-\Im(S_{21}S_{11}^*+S_{22}S_{12}^*)\\
Z_{42}&=-\Im(S_{21}S_{11}^*-S_{22}S_{12}^*)\\
Z_{43}&=\Im(S_{22}S_{11}^*-S_{12}S_{21}^*)\\
Z_{44}&=\Re(S_{22}S_{11}^*-S_{12}S_{21}^*)
\end{align}
where $\mathbf{S}$ is the standard $2\times 2$ amplitude scattering matrix linking incident and scattered transverse field vectors in the respective directions ($\hat{\mathbf{r}}^{\text {inc}}$) and ($\hat{\mathbf{r}}^{\text {sca}}$),\cite{MishchenkoTL02}

\begin{equation}
\left[\begin{array}{l}
E_{\theta}^{\text {sca }}\left(\hat{\mathbf{r}}^{\text {sca}}\right) \\
E_{\varphi}^{\text {sca }}\left(\hat{\mathbf{r}}^{\text {sca}}\right)
\end{array}\right]=\frac{\exp \left(\mathrm{i} \km r\right)}{r} \mathbf{S}\left(\hat{\mathbf{r}}^{\text {sca}}, \hat{\mathbf{r}}^{\text {inc}}\right)\left[\begin{array}{l}
E_{0 \theta }^{\text {inc }} \\
E_{0 \varphi }^{\text {inc }}
\end{array}\right].
\end{equation}
The amplitude scattering matrix $\mathbf{S}$ is derived from the collective \tmatrix\ following Ref.~\citenum{MishchenkoTL02} (Eqs.~5.277--5.280).
\subsubsection{Differential scattering cross-section}
The differential scattering cross-section describes the angular distribution of the scattered light. It depends on the polarisation of the incident wave as well as the incidence and scattering directions, and is readily calculated from the Stokes phase matrix and incident Stokes vector\citep{MishchenkoTL02}
\begin{equation}
\begin{split}
\frac{dC_\mathrm{sca}}{d\Omega}&=\frac{1}{I_{\mathrm{inc}}}[Z_{11}(\hat{\br},\hat{\bn}_{\mathrm{inc}})I_{\mathrm{inc}}+Z_{12}(\hat{\br},\hat{\bn}_{\mathrm{inc}})Q_{\mathrm{inc}}\\&\quad+Z_{13}(\hat{\br},\hat{\bn}_{\mathrm{inc}})U_{\mathrm{inc}}+Z_{14}(\hat{\br},\hat{\bn}_{\mathrm{inc}})V_{\mathrm{inc}}].
\end{split}
\end{equation}
\subsection{Near-field quantities}
Solving the linear system of multiple-scattering equations provides the scattered field coefficients, from which we can compute the complex vector fields $\mathbf{E},\mathbf{B}$ everywhere in space, as well as derived quantities such as $|\mathbf{E}|^2,|\mathbf{E}|^4,$ or the local degree of optical chirality $\ldoc\propto\Im(\mathbf{E}^*\cdot\mathbf{B})$. If only specific directions of incidence are needed, the system may be solved directly, without inversion ($\keywd{Scheme}=0$, fastest). However, in some circumstances, such as the description of randomly-oriented clusters, we also seek orientation-averaged near-field quantities, requiring $\keywd{Scheme}>0$.

Near-field values are calculated in $\keywd{Mode} = 1$ in \TERMS, with \subroutine{mapNF} the main subroutine which receives input values and dispatches to other subroutines for the calculation of specific near-field quantities. 

\subsubsection{Orientation averaged local field intensity}
\label{sec:E2OA}
Following Ref.~\citenum{auger2008local}, the local field intensity expressed in terms of the scatterer-centred \tmatrices\ $\bT^{(i,j)}$ can be averaged over all possible directions of light incidence, yielding
%
\begin{equation}
\langle |\bE_\mathrm{tot}(\it{k}\br)|^2\rangle = 2\pi E^2\left(A_0+B_0+C_0\right)
\end{equation}
where
\begin{equation}
A_0=1/2\pi,
\end{equation}
\begin{equation}
B_0=2 \Re \sum_{j=1}^{N}\sum_{l=1}^{N}\tr \Bigg(\rbW^{\dagger}(\br_l)\bP(\hat{\br}_l,\hat{\br}_j)\bW(\br_j)\bT_{N}^{(j,l)} \Bigg)   ,
\end{equation}
\begin{align}
C_0=\tr \Bigg(&\sum_{j=1}^{N}\sum_{l=1}^{N}\sum_{i=1}^{N}\sum_{k=1}^{N}\bO^{(l,k)}\bT_{N}^{\dagger(i,k)} {\bW}^{\dagger}(\br_i)\\&\bP(\hat{\br}_i,\hat{\br}_j){\bW}(\br_j)\bT_{N}^{(j,l)}   \Bigg)\nonumber
\end{align}
where $\bP(\hat{\br}_i,\hat{\br}_j)=C^t(\hat{\br}_i)C(\hat{\br}_j)$ and $C(\hat{\br}_j)$ transforms vector in the jth particle spherical coordinate basis to the cartesian coordinate system:
\begin{equation}
C(\hat{\br}_j)=
\begin{bmatrix}
\sin\theta_j\cos\varphi_j & \cos\theta_j  \cos\varphi_j & -\sin\varphi_j\\
\sin\theta_j\sin\varphi_j & \cos\theta_j  \sin\varphi_j &  \cos\varphi_j \\
\cos\theta_j & -\sin\theta_j & 0\\
\end{bmatrix}.
\end{equation}
The terms $A_0$, $B_0$, $C_0$ correspond to the incident electric field, the interference between incident and scattered electric field, and the scattered electric field, respectively.

The orientation average of the local field intensity is mainly calculated in the subroutine \subroutine{calcOaExtField} of the \module{multiscat} module.

%
\subsubsection{Optical chirality ($\ldoc$)}
In order to evaluate the local degree of optical chirality ($\ldoc$), the total electric $\bE=\bE_\mathrm{sca}+\bE_\mathrm{inc}$ and magnetic $\bB=\bB_\mathrm{sca}+\bB_\mathrm{inc}$ vector fields are first evaluated at the requested position, from which $\ldoc$ is obtained as \citep{Hentschel2013Complex2}
\begin{equation}
\ldoc =\frac{-\omega\varepsilon_0}{2}\Im(\bE^* .\bB)
\end{equation}
From the Maxwell equation $\bB = -i\omega^{-1} \nabla \times \bE$, the magnetic field is expressed in the VSWF basis with the same coefficients as the electric field (with a simple swap and a prefactor),
\begin{align}
 \bB(\br;\it{k})=\frac{-ik}{\omega}\sum_{n=1}^{\nmax}\sum_{m=-n}^{n}&[\ba_{1,nm}\bN_{nm}(\it{k}\br)+\\&\ba_{2,nm}\bM_{nm}(\it{k}\br)] \nonumber
\end{align}
Thus, the field coefficients ($a_{1,nm},a_{2,nm}$) provide us with both the electric and magnetic field, from which we derive the local degree of optical chirality $\ldoc$. The subroutine \subroutine{calcLDOC} of the \module{multiscat} module calculates $\ldoc$. The value of $\ldoc$ is often normalised with respect to the degree of chirality of circularly-polarised plane waves $\ldoc=\pm kE^2\varepsilon_0/2$ ($+$ for RCP and $-$ for LCP, respectively), with incident electric field $E \equiv |\bE_\mathrm{inc}|$, defining
\begin{equation}
\ldocbar = \frac{2}{k\varepsilon_0 E^2} \ldoc.
\end{equation}
%

%
\subsubsection{Orientation-averaged optical chirality $\ldocoabar$}
For the calculation of orientation-averaged local degree of optical chirality $\ldocoabar$, we combine the near-field averaging procedure of Sec.~\ref{sec:E2OA} with the treatment of optical activity in Sec.~\ref{sec:CD}, expressing electric and magnetic fields as VSWFs in the helicity basis. We refer the reader to Ref.~\citenum{fazel2021orientation} for details of the derivation, and summarise the result:
\begin{equation}
\langle \ldoc \rangle=2\pi k \varepsilon_{0}E^2\,\Re \left(A_0+B_0+C_0+D_0\right)
\label{eq:avg}
\end{equation}
where, for right-handed circular polarisation
\begin{equation}
\begin{split}
	A^\text{(R)}_0= & {-1}/{4\pi}    \\
	B^\text{(R)}_0= & \sum_{j=1}^{N}\sum_{l=1}^{N}\tr\left( %
	\bheltBr^\dagger(k\br_l)\left[-\bUr_{j,l} + \bVr_{j,l} \right] \right)\\
	C^\text{(R)}_0= & \sum_{j=1}^{N}\sum_{l=1}^{N}\tr\left( %
	\left[-\bUhr_{j,l} - \bVhr_{j,l} \right]\bheltBr(k\br_l)\right)  \\
D^\text{(R)}_0= & \sum_{j=1}^{N}\sum_{l=1}^{N}\sum_{i=1}^{N}\sum_{k=1}^{N}\tr \bigg( %
\bO_{\tinyRR}^{(k,l)}\left[\bUhr_{j,l} + \bVhr_{j,l}\right]\\
{} & \phantom{\sum_{j=1}^{N}\sum_{l=1}^{N}\sum_{i=1}^{N}\sum_{k=1}^{N}\tr \bigg( %
\bO_{\tinyRR}^{(k,l)}}\left[-\bUr_{i,k} + \bVr_{i,k}\right]\bigg).
\end{split} \label{eq:finalR}
\end{equation}
where $\bheltBr$ and $\bheltBl$ are the left and right regular VSWFs in helicity basis and ${\bhelB}_{\tinyR}$, ${\bhelB}_{\tinyL}$ the corresponding irregular VSWFs (cf Eqs.~\ref{eq:R-helicity basis}, \ref{eq:L-helicity basis}). The terms  $\bUr_{j,l}$ and $\bVr_{j,l}$ (and their Hermitian transpose) are introduced for conciseness and defined as,
\begin{equation}
\begin{aligned}
\bUr_{j,l}  &:= & \bhelBr(k\br_j)\bT_{\tinyRR}^{(j,l)} ;\quad%
\bUhr_{j,l} & = & \bT_{\tinyRR}^{\dagger (j,l)}\bhelBr^\dagger(k\br_j)\\
\bVr_{j,l}  &:= & \bhelBl(k\br_j)\bT_{\tinyLR}^{(j,l)} ;\quad%
\bVhr_{j,l} & =&  \bT_{\tinyLR}^{\dagger (j,l)}\bhelBl^\dagger(k\br_j)\\
\end{aligned}
\end{equation}
The corresponding formulas for left-handed circular polarisation read
\begin{equation}
\begin{split}
	A^\text{(L)}_0= & {+1}/{4\pi}    \\
	B^\text{(L)}_0= & \sum_{j=1}^{N}\sum_{l=1}^{N}\tr\left( %
	\bheltBl^\dagger(k\br_l)\left[\bUl_{j,l} - \bVl_{j,l} \right] \right)\\
	C^\text{(L)}_0= & \sum_{j=1}^{N}\sum_{l=1}^{N}\tr\left( %
	\left[\bUhl_{j,l} + \bVhl_{j,l} \right]\bheltBl(k\br_l)\right)  \\
D^\text{(L)}_0= & \sum_{j=1}^{N}\sum_{l=1}^{N}\sum_{i=1}^{N}\sum_{k=1}^{N}\tr \bigg( %
\bO_{\tinyLL}^{(k,l)}\left[\bUhl_{j,l} + \bVhl_{j,l}\right]\\
{} & \phantom{\sum_{j=1}^{N}\sum_{l=1}^{N}\sum_{i=1}^{N}\sum_{k=1}^{N}\tr \bigg( %
\bO_{\tinyLL}^{(k,l)}}\left[\bUl_{i,k} - \bVl_{i,k}\right]\bigg).
\end{split} \label{eq:finalL}
\end{equation}
with
\begin{equation}
\begin{aligned}
\bUl_{j,l} &:= & \bhelBl(k\br_j)\bT_{\tinyLL}^{(j,l)}  ;\quad%
\bUhl_{j,l}& =&  \bT_{\tinyLL}^{\dagger (j,l)}\bhelBl^\dagger(k\br_j)\\
\bVl_{j,l} &:= & \bhelB_{\tinyR}(k\br_j)\bT_{\tinyRL}^{(j,l)} ;\quad%
\bVhl_{j,l}& = & \bT_{\tinyRL}^{\dagger (j,l)}\bhelBr^\dagger(k\br_j)
\end{aligned}
\end{equation}
Note that the sum of $B_0$ and $C_0$ simplifies to,
\begin{equation}
\Re\left(B^\text{(R)}_0+C^\text{(R)}_0\right) = 
 -2\Re \left(\sum_{j=1}^{N}\sum_{l=1}^{N}\tr\left(%
\bheltBr^\dagger(k\br_l)\bUr_{j,l}\right)\right)
\end{equation}
and
\begin{equation}
\Re\left(B^\text{(L)}_0+C^\text{(L)}_0\right) = 
 2\Re \left(\sum_{j=1}^{N}\sum_{l=1}^{N}\tr\left(%
\bheltBl^\dagger(k\br_l)\bUl_{j,l}\right)\right).
\label{eq:BCL}
\end{equation}
These formulas (\ref{eq:avg}--\ref{eq:BCL}) are implemented in the subroutine \subroutine{calcOaLDOC} of the \module{multiscat} module. 


%
\subsection{Solution schemes}
\TERMS\ offers several options, selected by \keywd{Scheme}, to solve the multiple scattering problem described in Section~\ref{sec:STM}. It generally requires determining the particle-centred coefficients $\bc_{j}^{(j)}$ for given individual scatterer properties (described by $\bT_{1}^{(j)}$) and an excitation field (described by $\rba^{(j)}$) impinging from a particular direction. This can be achieved by solving the linear system of equations in (\ref{eqn:FunMultiScatMat}) for $\bc^{(j)}_{j}$ \emph{without} performing matrix inversion, thus producing a complete description of the scattered field for the given excitation. The linear system can be solved with multiple right-hand sides, representing different excitations, with standard linear algebra routines. Performing matrix inversion to determine the collective \tmatrix\ becomes worthwhile only when many impinging directions are to be considered, or when orientation-averaged quantities are of primary interest.

A brute force approach to solving the multiple scattering problem would be to construct the $N\lmax \times N\lmax$ matrix in equation (\ref{eqn:FunMultiScatMat}) and then invert it to obtain the pairwise scatterer-centred \tmatrices\ $\bT^{(i,j)}$. However, this approach is computationally demanding. To help alleviate the cost of one large matrix inversion, Stout~\emph{et al.}\cite{StoutAL02,StoutAD08} proposed a recursive scheme where a smaller ($\lmax \times \lmax$) matrix is inverted $N-1$ times.

\vfill\null
\subsubsection{Recursive scheme with matrix balancing}
In the recursive algorithm described by Stout~\emph{et al.}\cite{StoutAL02}, the elements of $\bT_{N}^{(j,i)}$ are accumulated recursively from auxiliary subsystems, incrementally built up from one to $N$ particles. The recursive system is prescribed by the following four equations:
%
\begin{align}
\bT_{N}^{(N,N)}  = & \Bigg[ \left[\bT_{1}^{(N)}\right]^{-1} \\
&\quad - \sum_{j=1}^{N-1} \bO^{(N,j)} \left( \underline{ \sum_{i=1}^{N-1} \bT_{N-1}^{(j,i)} \bO^{(i,N)} } \right) \Bigg]^{-1}\nonumber\\
 {} & =:  \mathbf{S}^{-1},  \label{eqn:TNNN}\\
\bT_{N}^{(N,i)}  = & \bT_{N}^{(N,N)} \left( \sum_{j=1}^{N-1} \bO^{(N,j)} \bT_{N-1}^{(j,i)} \right), \quad i\neq N,\\
\bT_{N}^{(j,N)}  = & \left( \underline{ \sum_{i=1}^{N-1} \bT_{N-1}^{(j,i)} \bO^{(i,N)} } \right) \bT_{N}^{(N,N)}, \quad j\neq N,\\
\bT_{N}^{(j,i)}  = & \bT_{N-1}^{(j,i)}  \\
&  + \left( \underline{ \sum_{l=1}^{N-1} \bT_{N-1}^{(j,l)} \bO^{(l,N)} } \right) \bT_{N}^{(N,i)}, \quad j,i\neq N, 
\end{align}
%
where a common matrix sum has been underlined. Note that only one $\lmax\times \lmax$ matrix is inverted on each of the $N-1$ iterations. The inverted matrix becomes ill-conditioned for large $\nmax$, but the associated problems can be (at least partly) circumvented by applying the recursive scheme to appropriately "balanced" matrices and coefficients:\cite{StoutAD08}
\begin{align}
\left[ \widehat{\bT}_{N}^{(j,i)} \right]_{sp,s'p'} := & \left[ \bD^{(j)} \bT_{N}^{(j,i)} [\rbD^{(i)}]^{-1} \right]_{sp,s'p'}  \\
 {} = &  \frac{\xi_{n(p)}(k_{\rm M}R_{j})}{\psi_{n'(p')}(k_{\rm M}R_{i})} \left[ \bT_{N}^{(j,i)} \right]_{sp,s'p'} \nonumber \\
\left[ \widehat{\bT}_{1}^{(j)} \right]_{sp,s'p'}  := & \left[ \bD^{(j)} \bT_{1}^{(j)} [\rbD^{(j)}]^{-1} \right]_{sp,s'p'}\\
 {} = &  \frac{\xi_{n(p)}(k_{\rm M}R_{j})}{\psi_{n'(p')}(k_{\rm M}R_{j})} \left[ \bT_{1}^{(j)} \right]_{sp,s'p'} \nonumber \\
\left[ \widehat{\bO}^{(j,i)} \right]_{sp,s'p'}  := & \left[ \rbD^{(j)} \bO^{(j,i)} [\bD^{(i)}]^{-1} \right]_{sp,s'p'} \\
 {} = & \frac{\psi_{n(p)}(k_{\rm M}R_{j})}{\xi_{n'(p')}(k_{\rm M}R_{i})} \left[ \bO^{(j,i)} \right]_{sp,s'p'}, \nonumber
\end{align}
where $\rbD^{(j)}$ and $\bD^{(j)}$ are regular and irregular diagonal matrices with the Riccati-Bessel functions $\psi_{n}(x) = x j_{n}(x)$ or $\xi_{n}(x) = x h_{n}(x)$ on the diagonal. (Here, $j_{n}(x)$ and $h_{n}(x)$ are, respectively, the spherical Bessel and Hankel functions of the first kind)~\cite{LeRuE08}. Note that Stout~\emph{et al.}~\cite{StoutAD08}'s "matrix balancing" may also be regarded as "left and right preconditioning", as the matrix to be inverted is essentially left- and right-multiplied by two different matrices to improve its condition number, which aims to make numerical inversion more robust and accurate. Instead of balancing throughout, as Stout~\emph{et al.}~\cite{StoutAD08} propose to do, we prefer to localise the balancing act just at the inversion step in (\ref{eqn:TNNN}), i.e.
\begin{align}
\bT_{N}^{(N,N)} = &\mathbf{S}^{-1} \nonumber\\
 = & \left[ \bD^{(N)} \right]^{-1} \bD^{(N)} \mathbf{S}^{-1} \left[ \rbD^{(N)} \right] ^{-1}\rbD^{(N)}  \\
 = & \left[ \bD^{(N)} \right]^{-1} \left[ \rbD^{(N)} \mathbf{S} \left[ \bD^{(N)} \right]^{-1} \right]^{-1}\rbD^{(N)} \nonumber \\
 =: & \left[ \bD^{(N)} \right]^{-1}  \widehat{\mathbf{S}}^{-1} \rbD^{(N)} 
\end{align}
where $\widehat{\mathbf{S}}$ is obtained by balancing $\mathbf{S}$ analogously to $\widehat{\bT}_{N}^{(j,i)}$. Note that equation (\ref{eqn:TNNN}) can be factored in two ways:
\begin{equation} \label{eqn:TNNN2}
\bT_{N}^{(N,N)}  =  \bT_{1}^{(N)} \mathbf{S}^{-1}_{R}  =  \mathbf{S}^{-1}_{L} \bT_{1}^{(N)} 
\end{equation}
where
\begin{align} \label{eqn:TNNN2b}
\mathbf{S}_{R} =& \left[ \bI - \sum_{j=1}^{N-1} \bO^{(N,j)} \left( \underline{ \sum_{i=1}^{N-1} \bT_{N-1}^{(j,i)} \bO^{(i,N)} } \right) \bT_{1}^{(N)} \right] \\
\mathbf{S}_{L}= & \left[ \bI - \bT_{1}^{(N)} \sum_{j=1}^{N-1} \bO^{(N,j)} \left( \underline{ \sum_{i=1}^{N-1} \bT_{N-1}^{(j,i)} \bO^{(i,N)} } \right) \right]
\end{align}
%
either of which may be preferred if $\bT_{1}^{(N)}$ is difficult to invert. To facilitate the inversion of $\mathbf{S}_{L}$ and $\mathbf{S}_{R}$, slightly different balancing (and subsequent unbalancing) should be used: 
\begin{align}
\bS_{L}^{-1} = & \left[\bD_{N}^{(N)}\right]^{-1} \left[ \bD_{N}^{(N)} \bS_{L} \left[\bD_{N}^{(N)}\right]^{-1} \right]^{-1} \bD_{N}^{(N)}\\
=: & \left[\bD_{N}^{(N)}\right]^{-1} \widehat{\bS}_{L}^{-1} \bD_{N}^{(N)}, \\
\bS_{R}^{-1} = & \widetilde{\bD}_{N}^{(N)} \left[ \left[\widetilde{\bD}_{N}^{(N)}\right]^{-1} \bS_{R} \widetilde{\bD}_{N}^{(N)} \right]^{-1} \left[\widetilde{\bD}_{N}^{(N)}\right]^{-1} \\
=: & \widetilde{\bD}_{N}^{(N)}  \widehat{\bS}_{R}^{-1} \left[\widetilde{\bD}_{N}^{(N)}\right]^{-1}.
\end{align}
Here $\bS_{L}$ is balanced 
using only irregular weights, while $\bS_{R}$ is balanced like a \tmatrix\ using only regular weights. In our experience, $\widehat{\bS}_{R}$ is much better conditioned for inversion than $\widehat{\bS}$.

In \TERMS\ program the balancing formulas are implemented in the subroutines \subroutine{balanceVecJ} and \subroutine{balanceMatJI}. 
\clearpage
\subsubsection{Mackowski \& Mishchenko's scheme}
From equations (\ref{eqn:TjiMatrixDefinition}) and (\ref{eqn:TjiMatrixFactored2}) it follows that 
\begin{widetext}
\begin{equation}
\begin{bmatrix}
\bI & \cdots & - \bT_{1}^{(1)} \bO^{(1,N)} \\
\vdots & \ddots & \vdots \\
- \bT_{1}^{(N)} \bO^{(N,1)} & \cdots & \bI
\end{bmatrix}\begin{bmatrix}
\bT_{N}^{(1,1)} & \cdots & \bT_{N}^{(1,N)} \\
\vdots & \ddots & \vdots \\
\bT_{N}^{(N,1)} & \cdots & \bT_{N}^{(N,N)}
\end{bmatrix}
=
\begin{bmatrix}
\bT_{1}^{(1)} & \cdots & 0 \\
\vdots & \ddots & \vdots \\
0 & \cdots & \bT_{1}^{(N)}
\end{bmatrix},
\end{equation}
\end{widetext}
or, equivalently,
\begin{equation}
\bT_{N}^{(i,i)} - \sum_{j\neq i}\bT_{1}^{(i)} \bO^{(i,j)}\bT_{N}^{(j,i)} = \bT_{1}^{(i)},
\end{equation}
which is a linear system of the general form $\mathbf{A}\mathbf{X} = \mathbf{B}$, where we want to find matrix $\mathbf{X}$ for a given $\mathbf{A}$ and $\mathbf{B}$, which contains multiple right hand sides. That is, each column of $\mathbf{X}$ and $\mathbf{B}$ can be treated independently, so we have to solve many linear systems of the form  $\mathbf{A}\mathbf{x}_{\nu} = \mathbf{b}_{\nu}$, where $\mathbf{x}_{\nu}$ and $\mathbf{b}_{\nu}$ are the $\nu$'th column of $\mathbf{X}$ and $\mathbf{B}$, respectively.

Mackowski \& Mishchenko (M{\&}M)~\cite{Mackowski96, MackowskiM11} "contract" the second particle index of $\bT_{N}^{(j,i)}$, using $\bT_{N}^{(j)} = \sum_{i} \bT_{N}^{(j,i)} \rbO^{(i,0)}$, to rewrite the linear system in terms of \emph{individual} (as opposed to pairwise) scatterer-centred \tmatrices\ $\bT_{N}^{(j)}$, i.e.
\begin{widetext}
\begin{equation} \label{eqn:Mackowski}
\begin{bmatrix}
\bI & \cdots & - \bT_{1}^{(1)} \bO^{(1,N)} \\
\vdots & \ddots & \vdots \\
- \bT_{1}^{(N)} \bO^{(N,1)} & \cdots & \bI
\end{bmatrix}
\begin{pmatrix}
\mathbf{T}_{N}^{(1)} \\
\vdots \\
\mathbf{T}_{N}^{(N)}
\end{pmatrix}
=
\begin{pmatrix}
\mathbf{T}_{1}^{(1)} \rbO^{(1,0)} \\
\vdots \\
\mathbf{T}_{1}^{(N)} \rbO^{(N,0)}
\end{pmatrix},
\end{equation} 
or, equivalently,
\begin{equation}
\bT_{N}^{(i)} - \sum_{j\neq i}\bT_{1}^{(i)} \bO^{(i,j)}\bT_{N}^{(j)} = \bT_{1}^{(i)} \rbO^{(i,0)},
\end{equation}
\end{widetext}
where the number of independent linear systems of the form $\mathbf{A}\mathbf{x}_{\nu} = \mathbf{b}_{\nu}$ is now reduced. M{\&}M use the biconjugate gradient method to solve $\mathbf{A}\mathbf{x}_{\nu} = \mathbf{b}_{\nu}$ for each $\nu$, where the row order (i.e.~length of each column vector) is predetermined from Mie theory for each (spherical) scatterer in isolation, and the truncation limit for the column order (i.e.~the maximum value of $\nu$) is determined on-the-fly from the convergence of each scatterer's contribution to the collective rotationally-averaged extinction cross-section (see Eq.~66 in Ref.~\citenum{Mackowski96}).

Mackowski \& Mishchenko's scheme is implemented as $\keywd{Scheme}= 3$ in \TERMS, with the addition of balancing discussed above, though we use a direct linear solver instead of an iterative one.

%% file: keywords.tex

\section*{Main input parameters}\label{main-input-parameters}

%
\vskip1em\noindent{\texttt{ModeAndScheme M S}}\\
If present, this keyword must appear first in the input file. It takes
two arguments: positive integer \emph{M} specifying the desired
calculation mode; and non-negative integer \emph{S} specifying the
solution scheme to be used. The default values are \emph{M} = 2 and
\emph{S} = 3.\\

\emph{Mode of calculation}

\begin{itemize}
\item
  \emph{M = 1} triggers a single- or multi-wavelength calculation of
  near fields \(\mathbf{E}\), \(\mathbf{B}\) and optical chirality
  \(\mathscr{C}\), at fixed incidence directions and/or
  orientation-averaged
\item
  \emph{M = 2} triggers a single- or multi-wavelength calculation of
  far-field properties (e.g.~spectra of optical cross-sections), at
  fixed incidence directions and/or orientation-averaged
\item
  \emph{M = 3} triggers a single- or multi-wavelength calculation of
  polarimetric properties, such as Stokes scattering vectors, phase
  matrix, and differential scattering cross sections for multiple
  incidence and/or scattering angles
\end{itemize}

\emph{Scheme of solution}

\begin{itemize}
\item
  \emph{S = 0} will seek solutions for the given angles of incidence, without
  seeking the collective \emph{T}-matrix
\item
  \emph{S \textgreater{} 0} will calculate the collective
  \emph{T}-matrix either (\emph{S = 1}) by direct inversion of the
  complete linear system to obtain $T^{(i,j)}$, or (\emph{S = 2}) by
  using Stout's iterative scheme for $T^{(i,j)}$, or (\emph{S = 3}) by
  using Mackowski \& Mishchenko's scheme for $T^{(i)}$. Note that fixed-orientation cross-sections are also
  calculated when $S > 0$.
\end{itemize}

\vskip1em\noindent{\texttt{Scatterers N}}\\
This keyword must appear last in the \texttt{inputfile}, with a single
positive integer argument \emph{N} specifying the number of scatterers.
The following \emph{N} lines specify all the required parameters per
scatterer, and each line must contain five or more space-separated
fields, i.e.

\begin{verbatim}
Tag x y z R [ a b c [ d ] ]     ( if Tag(1:2)  = "TF" )   
            [ a [ b [ c ] ] ]   ( if Tag(1:2) != "TF" )
\end{verbatim}

where \emph{Tag} is a contiguous string, which may contain one
underscore to separate the root from the suffix; \emph{x}, \emph{y},
\emph{z} are the cartesian coordinates (in the lab frame) for the scatterer, whose smallest circumscribing sphere has radius \emph{R}. All other subsequent parameters (inside square brackets)
depend on the root of \emph{Tag}.

Before the root of \emph{Tag} is parsed, the code first looks for a
suffix of the form \emph{\_S?} and associates it with a multipole
selection predefined using the
\protect\hyperlink{MultipoleSelections}{\texttt{MultipoleSelections}}
keyword.

If the root of \emph{Tag} is either ``TF1'', ``TF2'', \ldots{} , or
``TF9'', which correspond to a 1-body \emph{T}-matrix file listed under
the \protect\hyperlink{TmatrixFiles}{\texttt{TmatrixFiles}} keyword,
floats \emph{a}, \emph{b}, and \emph{c} can be supplied to specify the
Euler angles describing the scatterer orientation (default angle values
are all zero). Another float \emph{d} may be included to specify the
aspect ratio for spheroids, which is currently only used for mapping
local field intensity and visualising the geometry. Note that \emph{d}
is interpreted as the ratio of polar (along z) to equatorial (along x or
y) length, so that \emph{d \textgreater{} 1} for prolate spheroids,
\emph{d \textless{} 1} for oblate spheroids, and \emph{d = 1} for
spheres (default). Note that for nonspherical particles the
circumscribing sphere radius \emph{R} is used to check for potential
geometrical overlap between particles, but also in the balancing scheme.

If the root of \emph{Tag} is not ``TF?'', the 1-body \emph{T}-matrix is
computed using Mie theory, which is applicable to coated spheres. The
expected \emph{Tag} format is \verb+L0@L1@L2@L3+, with
the character ``@'' delimiting substrings that specify the material
dielectric function of each concentric region inside the scatterer,
starting from the core (\emph{L0}) and going \emph{outward}. The number
of coats is inferred from the number of instances of ``@'' and is
currently capped at 3. \emph{Tag} of a homogeneous sphere (without
layers) should not contain any ``@'', i.e.~\emph{Tag = L0}. Currently
accepted values for \emph{L?} are: ``Au'', ``Ag'', ``Al'', ``Cr'',
``Pt'', ``Pd'', ``Si'', and ``Water'' which trigger internal calculation
of the wavelength-dependent dielectric functions for the required
material, or ``DF1'', ``DF2'', \ldots, ``DF9'' to impose a custom
dielectric function listed under the
\protect\hyperlink{dielectricfunctions-nfuns-1}{\texttt{DielectricFunctions}}
keyword. For coated spheres, the outer radius of each region must be
specified by floats \emph{R}, \emph{a}, \emph{b}, \emph{c} in the order
of decreasing size (i.e.~going radially \emph{inward}).

\vskip1em\noindent{\texttt{TmatrixFiles Nfiles}}\\
Specifies the number of external \emph{T}-matrix files (default:
\emph{Nfiles = 0}). The subsequent \emph{Nfiles} lines are each read as
a string and then interpreted as a filename. Wrap the string in
quotation marks if it contains the relative path or special characters,
e.g.~\texttt{"../../tmatrix\_Au\_spheroid\_50x20\_water.tmat"}. Note
that the wavelengths in each file must \emph{exactly} correspond to the
values specified by the
\protect\hyperlink{wavelength-l1-l2-n-}{\texttt{Wavelength}} keyword.

The \tmatrix\ file format is as follows:

\begin{itemize}
\item
  First line is a comment (starts with a \texttt{\#}) describing the
  format \texttt{\#\ s\ sp\ n\ np\ m\ mp\ Tr\ Ti}
\item
  Second line is also a comment and starts with
  \texttt{\#\ lambda=\ N1\ nelements=\ N2} where N1 is the wavelength in
  nanometres, and N2 is the number of \emph{T}-matrix elements to be
  read below
\item
  Subsequent lines contain the indices and \emph{T}-matrix values for
  this particular wavelength,
\end{itemize}

\begin{enumerate}
\def\labelenumi{\arabic{enumi}.}
\item
  \texttt{s}, \texttt{sp} are the row (resp. column) index of the
  multipole mode (1: magnetic, or 2: electric)
\item
  \texttt{n}, \texttt{np} index the multipole degree
\item
  \texttt{m}, \texttt{mp} index the multipole order
\item
  \texttt{Tr}, \texttt{Ti} give the real and imaginary part of the
  \emph{T}-matrix element
\end{enumerate}

\begin{itemize}
\item
  If the file contains multiple wavelengths each wavelength-block is
  appended below the others, starting with a line
  \texttt{\#\ lambda=\ N1\ nelements=\ N2}
\end{itemize}

An example is show below,

\begin{verbatim}
# s sp n np m mp Tr Ti | prolate Au spheroid in water, a = 10 c = 20
# lambda= 400 nelements= 136 epsIn= -1.649657+5.771763j
  1   1   1   1  -1  -1 -1.189109253832815e-04 -2.161746691903687e-05
  1   1   1   1   0   0 -5.597968829951113e-05 -3.444956295771378e-05
... [truncated]
  2   2   4   4   3   3 -3.794740062663782e-11 5.636725538124517e-11
  2   2   4   4   4   4 -1.113090425618089e-11 1.707927691863483e-11
# lambda= 402 nelements= 136 epsIn= -1.661947+5.778032j
  1   1   1   1  -1  -1 -1.160926707256971e-04 -2.119092055798298e-05
  1   1   1   1   0   0 -5.467319805259745e-05 -3.371696756234449e-05
... [truncated]
  2   2   4   4   3   3 -1.279170882307354e-15 1.378894188143029e-13
  2   2   4   4   4   4 -3.752182192799965e-16 4.101975575297762e-14
... [truncated]
# lambda= 800 nelements= 136 epsIn= -24.236565+1.458652j
  1   1   1   1  -1  -1 -7.146139984375531e-07 -1.120611667309835e-05
  1   1   1   1   0   0 -4.379156367712547e-07 -7.955074171282911e-06
... [truncated]
  2   2   4   4   3   3 -1.240958755455683e-15 1.346747233206165e-13
  2   2   4   4   4   4 -3.640885008022631e-16 4.006450678480949e-14
... [truncated]
\end{verbatim}

\vskip1em\noindent{\texttt{DielectricFunctions Nfuns}}\\
Specifies the number of custom dielectric functions (default:
\emph{Nfuns = 0}). The subsequent \emph{Nfuns} lines are each read as a
string and then interpreted as either (i) a filename with a relative
path or (ii) real and imaginary parts of a constant (i.e.~wavelength
independent) value. Wrap each string in quotation marks,
e.g.~\texttt{"../../epsAg.dat"} or \texttt{"2.25d0\ 0.0d0"}. The files
should be in three-column format: the wavelength in nm followed by the
real and imaginary parts of the relative dielectric function on each line. The
wavelength range in the file must fully contain the range specified by
the \protect\hyperlink{wavelength-l1-l2-n-}{\texttt{Wavelength}}
keyword, but the values need not correspond exactly as they will be
linearly interpolated.

\vskip1em\noindent{\texttt{Medium X}}\\
Sets the real-valued dielectric constant of the host medium (default
value is \emph{1.0}). If \emph{X \textless{} 0} then its magnitude is
interpreted as a refractive index (\emph{s}), from which the dielectric
constant is calculated as $X=s^2$.

\vskip1em\noindent{\texttt{Wavelength L1 [ L2 n ]}}\\
Without the optional arguments, this keyword changes the default
wavelength of 666.0 nm to a new value \emph{L}1. Including the optional
arguments will specify a closed interval {[} \emph{L}1, \emph{L}2 {]}
divided into \emph{n} regular grid spacings, thus producing \emph{n+1}
wavelengths.

\vskip1em\noindent{\texttt{Incidence a b c [ p ] / [ na nb nc ]}}\\
or\\
\noindent\texttt{Incidence\ file\ filename\ {[}p{]}}\\
This keyword modifies the incident plane-wave. The default travel
direction (along \emph{z} in lab-frame) can be changed by the Euler
angles \emph{a} in the range \([0,2\pi)\) and \emph{b} in the range
\([0,\pi]\), coinciding with the azimuthal and the polar angles,
respectively, of the spherical polar coordinates in the lab frame. In
addition, the amplitude vector can then be rotated about the new travel
direction by the third Euler angle \emph{c} in the range \([0,2\pi)\).
All three Euler angles are defined in accordance with the right-hand
rule, and the sequence of rotation angles \emph{a},\emph{b},\emph{c}
corresponds to the intrinsic ZY'Z' convention. That is: rotate by
\emph{a} about the current \emph{z}-axis, then by \emph{b} about the new
\emph{y}-axis, and finally by \emph{c} about the new \emph{z}-axis.

Near-field and polarimetric calculations, i.e.~in modes \emph{M = 1} and
\emph{M = 3}, require the polarisation of incident light to be
specified. The polarisation is set by integer \emph{p}, with
\emph{\textbar p\textbar{} = 1} setting linear polarisation,
\emph{\textbar p\textbar{} = 2} setting circular polarisation, and the
sign selecting one of the two Jones vectors in each case (positive:
\emph{x}-linear-polarised or \emph{R}-circular-polarised; negative:
\emph{y}-linear polarised or \emph{L}-circular-polarised). Note: for a
circularly polarised wave travelling along \emph{z}, right-circular
(\emph{R}) polarisation means that the amplitude vector is rotating
clockwise in the \emph{xy}-plane from the receiver's viewpoint (looking
in the negative \emph{z} direction).

The integer \emph{p} can be omitted in mode \emph{M = 2}, because its
output is always calculated for all four polarisations.

A negative value of argument \emph{a}, \emph{b}, and/or \emph{c} will
trigger discretisation of the corresponding angle range to produce
$-a$ grid points (resp. $-b$ or $-c$). The grid points are uniformly spaced for the
first and the third Euler angles, but for the second (i.e.~polar) angle
the discretisation is such that the cosine is uniformly spaced. Note that the discretisation is
constructed so that orientational averages are computed as a uniformly
weighted Riemann sum with the midpoint rule. The weight $w_i$ of
each grid point \emph{i} is simply $w_i = 1/n_\text{gps}$, where $n_\text{gps}$
is the total number of grid points. The range maximum of each angle can be divided by an (optional) integer \emph{na}, \emph{nb}, and \emph{nc}, to help avoid evaluating redundant
grid points in the presence of symmetry. 

Multiple incidences can also be read from a file, in which case the
argument \emph{a} must be a string starting with `f' or `F', and
\emph{b} must specify the filename. The file's first line must contain
the total incidence count, \emph{ninc}, and the subsequent \emph{ninc}
lines each must contain four space-separated values: the three Euler
angles (\emph{ai}, \emph{bi}, \emph{ci}) and the weight $w_i$ of
each incidence. The weights are only used to compute rotational
averages for convenience, which is a common use-case.

In \texttt{Mode\ =\ 1} (near-field calculations), if \texttt{p} is set
to \texttt{p=1} (default value, linear polarisaiton), the orientation
average of the local degree of optical chirality
\(\langle\mathscr{C}\rangle\) will be calculated for both RCP and LCP
(noting that linear polarisation would give 0 everywhere, when
orientation-averaged). Since the calculation can be time-consuming,
setting \texttt{p=+/-2} triggers the calculation for only that specific circular
polarisation.

\vskip1em\noindent{\texttt{MultipoleCutoff n1 [ n2 [ t ] ]}}\\
Change the primary multipole cutoff (used for irregular offsetting when
staging the linear system) from the default value of 8 to \emph{n1}.
Another cutoff (used for regular offsetting when ``contracting'' the
collective \tmatrix) can be set to \emph{n2} \textgreater= \emph{n1} (equality by
default). A relative tolerance $10^t$ (with $t<0$
and $t = -8$ by default) is used in the test for convergence of cross-sections with
respect to multipole order $n=1\dots n_2$ (the summation can terminate below $n_2$ if the relative tolerance is reached).

\vskip1em\noindent{\texttt{MultipoleSelections Ns}}\\
This keyword defines optional multipole selections for individual
\tmatrices, and it must be followed by \emph{Ns} lines with two
fields: (i) a string \emph{range} specifying the selection range; and
(ii) a string \emph{type} specifying the selection type. For example:

\begin{verbatim}
MultipoleSelections 3
MM1:4_EM1:4_ME1:4_EE1:4  blocks
MM1:0_EM1:15_ME1:8_EE1:0  rows
EM1:1_ME1:1  columns
\end{verbatim}

The \emph{range} string must be of the form MM?:?\_EM?:?\_ME?:?\_EE?:?,
with the underscores separating the ranges for each \emph{T}-matrix
block (e.g.~MM or ME), and each range specified by a closed multipole
interval ?:? (e.g.~\emph{n}lo:\emph{n}hi = 1:4). No selection will be
applied to blocks not included in \emph{range}, so these ``missing''
blocks will remain unmasked (left "as is" in the original \tmatrix). On the other hand, a whole block can be masked (zeroed out) by setting \emph{n}lo \textgreater{} \emph{n}hi (e.g. \verb+MM1:0+ will set the whole \verb+MM+ block of the \tmatrix\ to 0).

The \emph{type} string must either start from ``c'', ``r'', or ``b'', to
indicated that the selection is either applied to \emph{T}-matrix
columns, rows, or both (producing non-zero blocks). To clarify, if
\emph{type(1:1) = ``c''}, then all \emph{T}-matrix columns corresponding
to multipole orders \emph{n \textless{} n}lo and \emph{n \textgreater{}
n}hi will be set to zero. For \emph{type(1:1) = ``b''}, columns
\textbf{and} rows for \emph{n \textless{} n}lo and \emph{n
\textgreater{} n}hi will be set to zero.

\section*{Output control}\label{output-control}

%
\vskip1em\noindent{\texttt{OutputFormat F [ filename ]}}\\
If present, the output file format \emph{F} can be switched between plain text
(``TXT'', default) and HDF5 (``HDF5''). With ``HDF5'', the results will
be stored in a file with name ``results.h5'', or a user-specified
filename (extension \emph{.h5} added automatically).

\vskip1em\noindent{\texttt{Verbosity L}}\\
Keyword specifying integer-valued verbosity level \emph{L}. Silent mode
(\emph{L = 0}) prints only error statements and warnings. Physical
quantities and some status indicators are printed at low verbosity
(\emph{L = 1}, default value), with various timings and convergence
indicators released at medium verbosity (\emph{L = 2}). The highest
level (\emph{L = 3}) is intended for debugging, releasing all print
statements throughout the code.

\section*{Near-field specific
keywords}\label{near-field-specific-keywords}

%
\vskip1em\noindent{\texttt{SpacePoints filename}}\\
or\\
\noindent{\texttt{SpacePoints xlo xhi nx ylo yhi ny zlo zhi nz}}\\
Read (from a file) or calculate (on a regular grid) the cartesian
coordinates of points in space, where the local field quantities are to be
evaluated. The file's first line should contain the total number of
space-points, and the subsequent lines must contain the \emph{x},
\emph{y}, and \emph{z} coordinates of each point. A regular grid is
specified by a closed interval, e.g.~\emph{{[} xlo, xhi {]}}, and the
number of bins (\emph{nx}) the interval is to be divided into (thus
producing \emph{nx+1} grid points along that dimension).

\vskip1em\noindent{\texttt{MapQuantity [p] [E] [B] [C]}}\\
Specify the near-field quantities of interest, in \texttt{Mode\ =\ 1}. Integer argument \emph{p} selects the raising power applied to the field amplitude \(|\mathbf{E}|^p\) or \(|\mathbf{B}|^p\). The default is \emph{p = 2} yielding the field intensity, \emph{p = 1} is for the field
amplitude $|E|$, \emph{p = 4} for the (approximate) Raman enhancement
factor $|E|^4$. Setting \emph{p = 0} will output the real and imaginary parts of the (vector!) field
components at each space-point.

The optional letters {[}\emph{E}{]} {[}\emph{B}{]} {[}\emph{C}{]}
(default: \emph{E} only) determine which of the near-field properties
(electric and magnetic fields and normalised value of local degree of
optical chirality) will be calculated.

\vskip1em\noindent{\texttt{MapOaQuantity [E] [B] [C]}}\\
This is a keyword applicable in \texttt{Mode\ =\ 1}, to request the
calculation of analytical orientation-averaged near-field quantities
\(\langle|\mathbf{E}|^2\rangle\), \(\langle|\mathbf{B}|^2\rangle\), or \(\langle\overline{\mathscr{C}}\rangle\).
If this keyword is not included in the input file, by default none will
be calculated. Note that the
\protect\hyperlink{incidence-a-b-c-p-na-nb-nc-}{\texttt{Incidence}}
keyword is used to select LCP and RCP (or both).

\section*{Polarimetry keywords}\label{polarimetry-keywords}

%
\vskip1em\noindent{\texttt{ScatteringAngles  a  b  c  / [ na  nb  nc ]}}\\
This keyword specifies the scattering angles in \texttt{Mode\ =\ 3}
(polarimetry), for the calculation of Stokes scattering vectors at
different scattering angles. The parameters have the same interpretation as for
\protect\hyperlink{incidence-a-b-c-p-na-nb-nc-}{\texttt{Incidence}}.

Multiple scattering angles can also be read from a file, in which case
the argument \texttt{a} must be a string starting with `f' or `F', and
\texttt{b} must specify the filename. The file's first line must contain
the number of scattering angles, \emph{nsca}, and the subsequent
\emph{nsca} lines each must contain three space-separated values: the
three Euler angles (\emph{ai}, \emph{bi}, \emph{ci}) for each scattering
angle.

\section*{Advanced use / development}\label{advanced-use-development}

%
\vskip1em\noindent{\texttt{ScattererCentredCrossSections}}\\
Applicable in \keywd{Scheme} 1 and 2. Triggers Stout's formulae for fixed and orientation-averaged
cross-sections based on scatterer-centred matrices; otherwise, the
default behaviour is to collapse the coefficients to a common origin.
Note that this does not affect the calculation of fixed-orientation
partial shell absorptions for layered spheres, as they are calculated
separately.

\vskip1em\noindent{\texttt{DumpCollectiveTmatrix [ filename ]}}\\
If the collective \emph{T}-matrix is computed, this keyword will dump it to a
file ``tmat\_col.txt'' or a user-specified \emph{filename}. The
\protect\hyperlink{tmatrixfiles-nfiles-1}{file format} is
self-consistent, so that the generated \emph{T}-matrix can be fed back
into \terms\  for subsequent calculations.

\vskip1em\noindent{\texttt{DumpPrestagedA}}\\
If present, dumps a sparse-format representation of the full matrix
comprising the individual \tmatrices\  after potential masking
followed by rotation in their respective frame.

\vskip1em\noindent{\texttt{DumpStagedA}}\\
If present, dumps a sparse-format representation of the full matrix
comprising the individual \tmatrices\  along the diagonal blocks,
and translation matrices in the off-diagonal blocks. The exact form of
this matrix is scheme-dependent.

\vskip1em\noindent{\texttt{DumpScaCoeff}}\\
If present, dumps the scattering coefficients into a file "Sca\_coeff" for different incidence angles.

\vskip1em\noindent{\texttt{DumpIncCoeff}}\\
If present, dumps the incident coefficients in to a file "Inc\_coeff" for different incidence angles.

\vskip1em\noindent{\texttt{DisableStoutBalancing}}\\
If present, switches off the balancing.

\vskip1em\noindent{\texttt{DisableRTR}}\\
Switches off the three-step translation of \tmatrices, where a
general translation is decomposed into a rotation, \emph{z}-axial
translation, and then the inverse rotation. Instead, a one-step
transformation is performed by pre- or post-multiplying by a single
matrix containing the general translation-addition coefficients.

%% file: subroutines.tex

In this section we briefly describe the main program, its different modules, subroutines, and functions. The arguments of each subroutine or function are shown in parentheses with {\color{\out}output arguments} highlighted in colour (note that some of them are both input and output arguments).
\subsubsection*{Main program \module{termsProgram}}
\module{termsProgram} is the main module, with subroutines listed below; it reads the keywords and their corresponding values in the input file and then calls different subroutines of \module{multiscat} module for the calculation of requested outputs.
\begin{itemize}
\item \subroutine{readInputFile(inputfile)}\\
Reads the input file containing specific keywords and the corresponding parameter values.

\item \subroutine{errorParsingArguments(keyword)}\\ 
If there is an error with the parameter values assigned to a keyword, this subroutine informs the user and stops the program.

\item \subroutine{calcEpsilon()}\\
Updates the \texttt{escat} array storing the relative dielectric function(s) for each scatterer evaluated at the specified wavelengths. \\
\texttt{escat} is an array for which the number of rows, columns and the 3$^{rd}$ dimension correspond to the number of shells, scatterers, and wavelengths, respectively.

\item \subroutine{calcGridPoints({\color{\out}points})}\\
For a near-field calculation, this subroutine calculates the grid points based on the number of steps, lower and upper values along the desired axes.\\
\texttt{points(3, nGridPoints)} is an in/output matrix storing the cartesian coordinates $(x,y,z)$ of the grid points.

\item \subroutine{sentence2words({\color{\out}sentence, words, nwords\_})} \\
Reads each line of the input file as a sentence and splits it into space-separated words.\\
\texttt{sentence}, \texttt{words} are in/output character arrays, and \texttt{nwords\_} is an optional output integer containing the number of words in the sentence.

\item \subroutine{dumpNFs2TXTFile(filename, incidences, Epower, wavelen, work, Ef, p\_label)}\\
Exports electric and magnetic near fields into a plain text file. \texttt{filename} is the name of the text file. \texttt{incidences, Epower, wavelen, work} are the incidence angles, selected power for mapping fields, wavelength, and near-field quantities, respectively. \texttt{Ef} is a logical flag which selects either electric field or magnetic field. \texttt{p\_label} is an integer array indexing the position of each grid point, whether it is inside the surrounding medium or a particle, and in which layer. 

\item \subroutine{dumpNFs2HDF5File(fname, groupname, filename, incidences, Epower, wavelen, work, p\_label)}
Exports electric and magnetic near fields into a HDF5 file. \texttt{filename}, \texttt{fname}, \texttt{groupname} are the names of the HDF5 file, group, and subgroup name, respectively. \texttt{incidences, Epower, wavelen, work} are the incidence angles, selected power for mapping fields, wavelength, and near-field quantities, respectively. \texttt{Ef} is a logical flag which selects either electric field or magnetic field. \texttt{p\_label} is an integer array indexing the position of each grid point, whether it is inside the surrounding medium or a particle, and in which layer. 

\item \function{countLines(filename) result({\color{\out}nlines})}\\
Counts lines in a text file.
\end{itemize}
\subsubsection*{\module{multiscat} module}
This module consists of a mix of high-level, core, low-level and supplementary routines for solving a multiple scattering problem using the \tmatrix\ formalism. We list below the subroutines of the \module{multiscat} module with a brief explanation. A list of common arguments and their brief description is at the end of this section. The other arguments are explained after each subroutine. 
 \begin{itemize}                       
\item \subroutine{mapNF(ncut, wavelen, inc,ehost, geometry, scheme, tfiles\_, escat\_,  nselect\_, verb\_, noRTR\_, dump\_oaE2, dump\_oaB2,
{\color{\out}field, Bfield, N\_OC, orAvextEB\_int, oa\_ldoc, p\_label})}\\
Calculates the electric and magnetic near fields, and normalised optical chirality ($\ldocbar$) for a multiple scattering problem, for different incidence directions and wavelengths, as well as the orientation-averaged value of external $\langle|\bE|^2\rangle, \langle|\bB|^2\rangle$ and $\ldocoabar$.                                                 
\texttt{escat\_, tfiles\_, nselect\_, verb\_, noRTR\_} are optional inputs.                             
\texttt{dump\_oaE2, dump\_oaB2} are logical flags selecting whether the orientation-averaged values  $\langle|\bE|^2\rangle$ and $\langle|\bB|^2\rangle$ will be calculated, respectively.

\item \subroutine{spectrumFF(ncut, wavelen,  ehost, geometry, scheme, escat\_, tfiles\_, nselect\_, noRTR\_, verb\_, {\color{\out}sig\_oa\_, sig\_, sig\_abs\_, jsig\_abs\_oa})}\\
Calculates cross-section spectra for (multiple) fixed orientations, partial absorptions, and orientation-averaged cross-sections for a particle cluster. \tmatrices\  for individual scatterers are either constructed using Mie theory or read from an optional argument \texttt{tfiles\_}. \texttt{escat\_, tfiles\_, nselect\_, verb\_, noRTR\_} are optional inputs.                             
\texttt{jsig\_abs\_oa} contains the orientation-averaged absorption cross-section of each particle (valid for homogeneous spheres only, at present).
 
\item \subroutine{solve(wavelen, ehost, geometry, nselect\_, scheme\_, verb\_, noRTR\_, {\color{\out}TIJ, cJ\_, cJint\_, csAbs\_, ierr\_} )}\\
This routine is the crux of \TERMS\ and solves a given multiple scattering problem by operating in a specified scheme. \texttt{TIJ} is an in/output argument, \texttt{cJ\_, cJint\_, csAbs\_} are optional in/output arguments,  
\texttt{nselect\_, scheme\_,  verb\_, noRTR\_} are optional inputs, and \texttt{ierr\_} is an optional output.
\texttt{TIJ} ($l_{max}\times \text{nscat}$, $l_{max}\times\text{nscat}$) as the input argument stores the \tmatrix\ of nonspherical particles as the diagonal blocks of the matrix, or dielectric values of different shells for spherical particles as the diagonal elements of the matrix. \texttt{nscat} is the number of scatterers. This subroutine updates and returns \texttt{TIJ} for the whole system as the output. \texttt{cJ\_(nscat x $l_{max}$, 2, nfi)} as the input argument contains details of the incident field and as a output argument contains incident plane wave coefficients in the first column and scattering coefficients in the second column. \texttt{nfi} is the number of incident angles.
\texttt{cJint\_(nscat x $l_{max}$, 4, 2)}: contains the regular and irregular field coefficients for each concentric region inside spherical scatterers.
\texttt{csAbs\_(nscat,4)}: contains absorption cross section inside each shell of each spherical scatterer.

\item \subroutine{stageAmat(scatXYZ, scatMiet, rtr, right\_, balance\_, verb\_, {\color{\out}A, Tmats\_})}\\
Stages a pre-staged matrix $A$.\\
\texttt{A (\textbf{$l_{max}$} x nscat, \textbf{$l_{max}$} x nscat)}: an in/output matrix, which must contain 1-body \tmatrices\  in the diagonal blocks on input and is a pre-staged matrix on the output; \texttt{right\_, balance\_, verb\_} are optional inputs; \texttt{Tmats\_(\textbf{$l_{max}$}, \textbf{$l_{max}$}, nscat)}: an optional output matrix which contains the 1-body \tmatrix\ of each particle.  
\texttt{balance\_}: a logical input argument which determines whether balancing is applied or not. 

\item \subroutine{calcTIJStout(scatXYZ, scatMiet, rtr, {\color{\out}TIJ})}\\
Calculates the scatterer-centred \tmatrix\ using the recursive scheme presented in Refs.~\citenum{StoutAL02} and \citenum{StoutAD08}. The relevant equations are 33 and 35 in Ref.~\citenum{StoutAL02}, and 20, 22 and 24 in \citenum{StoutAD08}. \texttt{TIJ} is an in/output argument.
\texttt{TIJ($l_{max}$ x nscat, $l_{max}$ x nscat)} as the input argument stores the \tmatrix\  of nonspherical particles as the diagonal blocks, or dielectric values of different shells for spherical particles as diagonal blocks. 
 
\item \subroutine{calcTIMackowski(scatXYZ, scatMiet, rtr, {\color{\out}TIJ)}}\\
Calculates the cluster's \tmatrix\ using Mackowski \& Mishchenko's formulation. \texttt{TIJ} is an in/output argument.
\texttt{TIJ($l_{max}$ x nscat, $l_{max}$ x nscat)} as the input argument stores the \tmatrix\  of nonspherical particles as the diagonal blocks, or dielectric values of different shells for spherical particles as diagonal blocks. The output \texttt{TIJ} is the scatterer-centred \tmatrices\  calculated using Mackowski \& Mishchenko's scheme\cite{Mackowski94,Mackowski96, MackowskiM11}. 

\item \subroutine{balanceMatJI(j, jregt, iregt, i, rev\_, mnq\_, {\color{\out}Mat})}\\
Performs balancing on a matrix (\texttt{Mat}) using two weights (indexed by $j$ and $i$). \texttt{Mat} is here taken as relating two vectors of VSWF coefficients, $c_j$ (centred at $j$) and $c_i$ (centred at $i$), such that $c_j = \text{Mat}\, c_i$. Logical inputs \texttt{jregt} and \texttt{iregt} specify whether $c_j$ and $c_i$ are regular or not. \texttt{Mat} is an in/output argument.

\item \subroutine{balanceVecJ(j, jregt, rev\_, {\color{\out}Vec})}\\
Performs balancing on a single vector (\texttt{V}) with a weight indexed by $j$. \texttt{V} corresponds here to the VSWF coefficients of particle $j$. \texttt{Vec} is an unbalanced/ balanced vector as the in/output argument.  
\texttt{j} specifies the scatterer.

\item \subroutine{calcCsStout(scatXYZR, aJ, fJ, nmax2\_, tol\_, verb\_, {\color{\out}sig})}\\
Calculates the extinction, scattering and absorption cross-sections from the incident and scattered coefficients using the Stout formulae\cite{StoutAL02}. \texttt{nmax2\_, tol\_, verb\_} are optional inputs and \texttt{sig} is an in/output matrix.

\item \subroutine{calcCs(scatXYZR, inc, fJ, nmax2\_, tol\_, verb\_, {\color{\out}sig})}\\
Calculates the extinction, scattering and absorption cross-sections from the incident and scattered coefficients which are collapsed to the common origin. Depending on the dimension of the \texttt{sig}, each cross-section is either just a total sum, or resolved into contributions from the multipole orders. \texttt{inc}: a vector of incidence angles. 
    
\item \subroutine{calcOAprops(Tmat, rtol\_, {\color{\out}sigOA, verb\_})}\\
Calculates orientation-averaged cross-sections and circular dichroism (CD) by transforming the \tmatrix\  (\texttt{Tmat}) from "parity" (M--N) basis to "helicity" (L--R) basis, following Ref.~\citenum{Suryadharma18}. \texttt{rtol\_} is an optional input, \texttt{verb\_} is an optional output, and \texttt{sigOA} is an in/output matrix containing orientation-averaged cross-sections and CD in each column for $n=1, \dots, n_{max}$.  

\item \subroutine{contractTmat(Tin, scatXYZR, rtr, mack\_, {\color{\out}Tout, verb\_})}\\
Combines the scatterer-centred \tmatrices\  into a common origin; the output will be the collective \tmatrix\ (\texttt{Tout}). \texttt{verb\_} is an optional in/output, \texttt{mack\_} is an optional logical input to calculate the collective \tmatrix\ based on Mackowski \& Mishchenko's scheme \cite{Mackowski94,Mackowski96,MackowskiM11}. 

\item \subroutine{diagnoseTmat(mode\_, verb\_, {\color{\out}Tmat})}\\
Determines the value of $n \leq n_{max}$ when $\trace(\Re{(\text{Tcol})})$ converges to $\text{rtol\_G}:=10^{-\text{ncut(3)}}$. 
If mode\_ $> 0$, also tests for the general symmetry, which applies to all
\tmatrices. (See equation 5.34 on p. 121 of Mishchenko\cite{MishchenkoTL02}).
    
\item \subroutine{calcOaStout(TIJ, scatXYZ, verb\_, {\color{\out}sigOA, cdOA\_, jAbsOA})}\\
Calculates the orientation-averaged extinction and scattering cross-sections defined in equations 44 and 47 of Stout \cite{StoutAL02}. The absorption cross-section is then deduced as the difference.
\texttt{TIJ} is the collective \tmatrix, \texttt{sigOA(3)} contains orientation-averaged extinction and scattering cross-sections, and \texttt{cdOA\_} is an optional output containing the corresponding values of CD.   
\texttt{jAbsOA}: contains the orientation-averaged absorption cross-section for each particle.

\item \subroutine{applyRotTranzRotOnMat(vtacs, bigdOP, rightOP, {\color{\out}mat})}\\ 
Performs the factorised translation of \tmatrices\  when changing origin. Instead of a single multiplication of a \tmatrix\ by a dense matrix containing the general translation-addition coefficients, this routine executes three multiplications by sparse matrices representing 1) a rotation, 2) a translation along the z-axis; and 3) an inverse rotation. This is meant to be more efficient when high multipole orders are included.\\
\texttt{vtacs(2x pmax,2 x pmax)}: axial VTACs with $(m,n,q)$ indexing, \texttt{bigdOP(pmax, pmax)}: optional input for rotation, \texttt{rightOP}: an optional logical input argument for applying the product from the right. \texttt{mat}: a non/translated matrix as the in/output. 

\item \subroutine{calcField(r, geometry, ipwVec, ipwE0, scaCJ, intCJreg\_, intCJirr\_, scatK\_, verb\_, {\color{\out}reE, imE, reB, imB, reE\_sca,imE\_sca, reB\_sca,imB\_sca, p\_label})}\\ 
Calculates the electric and magnetic near-field values at the determined grid points.\\
\texttt{r}: a matrix containing the coordinates of the grid points; \texttt{ipwVec(3)},\texttt{ipwE0(3)}: contain the wavevector and amplitude of the incident field, respectively; \texttt{scaCJ}: a vector containing scattering coefficients, \texttt{intCJreg\_, intCJirr\_}: contain the regular and irregular parts of the incident field coefficients transformed to the origin of each particle, respectively, \texttt{scatK\_} is the wavenumber in the host medium, and  
\texttt{reE, imE, reB, imB, reE\_sca,imE\_sca, reB\_sca,imB\_sca}: contain real(re) and imaginary(im) parts of the total electric (E) and magnetic (B) fields and the scattered field values at the grid points. 

\item \subroutine{dumpTmat(tmat, filename, lambda, eps\_med, tol\_, verb\_)}\\
Routine for dumping the collective \tmatrix\ (\texttt{tmat}) to a file in the format: 
\begin{center}
\texttt{s, s', n, n', m, m', T\_re, T\_im}
\end{center}
\texttt{filename} is an argument of type character corresponding to the name of the output file; \texttt{lambda}: the value of wavelength; \texttt{eps\_med}: the dielectric value of the host medium. 
  
\item \subroutine{dumpMatrix(mat, ofile, tolOP, verb\_)}\\
Outputs matrix \texttt{mat} to a desired optional tolerance (\texttt{tolOP}). \texttt{ofile}: the name of the output file. 

\item \subroutine{offsetTmat(off, miet, rtr, right, bigD\_, useD\_, balJI\_, {\color{\out}Tmat})}\\
Offsets the supplied \tmatrix\ \texttt{Tmat} by \texttt{off}, which can be either a square matrix of VTACs or a (note: complex!) displacement vector $k\mathbf{r}$(3) from which VTACs will be generated. Regular or irregular VTACs will be generated depending on whether $k\mathbf{r}$(3) is purely real or purely imaginary. If the logical input \texttt{miet} is true, Tmat will be treated as diagonal. If the logical input \texttt{rtr} is true, then offsetting will be based on factorised translation. If the logical input \texttt{right} is true, then offsetting will be done by post-multiplying Tmat from the right. \texttt{balJI\_} triggers balancing of the VTACs and the \tmatrices\ individually, before offsetting, but currently works only without factorised translation.

\item \subroutine{readTmatFile(filename, unit, wavelen, verb\_, {\color{\out}Tmat})}\\
Reads a \tmatrix\ from the input file (\texttt{filename}) and import it into the matrix \texttt{Tmat}. \texttt{unit}: an integer indexing the name of the \tmatrix\  file. \texttt{wavelen} is the value of the wavelength.   

\item \subroutine{parseInc(inc, verb\_, {\color{\out}inc\_dirn, inc\_ampl})}\\
Calculates the amplitude and direction vector of the incident plane wave based on the input Euler angles ($\alpha, \beta, \gamma$).
\texttt{inc\_dirn} and \texttt{inc\_ampl} are vectors containing the wavevector and amplitude of the incident electric field in cartesian coordinates, respectively. 
\texttt{inc} is a vector consisting of polarisation type and Euler angles of the incidence direction.

\item \subroutine{calcStokesScaVec(sca\_angles, inc2, ncut, wavelen, ehost, geometry, scheme, tfiles\_, escat\_, nselect\_, noRTR\_, verb\_, {\color{\out}StokesPhaseMat, StokesScaVec, diff\_sca})}\\
Calculates the Stokes phase matrix (\texttt{StokesPhaseMat}), Stokes scattering vector (\texttt{StokesScaVec}), and differential scattering cross-sections (\texttt{diff\_sca}).\\
\texttt{sca\_angles} is a matrix of desired scattering angles; if it is not specified in the input file, they are taken equal to the incidence angles. \texttt{inc2} is a matrix containing incidence angles. 

 \item \subroutine{calcLDOC(Ef, Bf, verb\_, {\color{\out}N\_OpC})} \\
Calculates the normalised optical chirality ($\ldocbar$) relative to the optical chirality of circularly polarised light. \texttt{Ef}, \texttt{Bf}, and \texttt{N\_OpC} are matrices containing the electric and magnetic field, and $\ldocbar$ values at the grid points, respectively.\\ 

 \item \subroutine{calcOaExtField(r, geometry, TIJ, lambda, ehost, escat, p\_label, verb\_, {\color{\out}orEB2})} \\ 
Calculates the orientation average of the total external electric and magnetic field intensities.
\texttt{r} is a matrix containing the cartesian coordinates of the grid points. \texttt{TIJ} is the scatterer-centred \tmatrix\ of the cluster. \texttt{orEB2} is a vector containing the value of orientation-averaged external electric and magnetic field intensities at the grid points.
\item \subroutine{calcOaLDOC(pol\_type, r, geometry, TIJ, verb\_, {\color{\out}Or\_OC})}\\
Calculates the orientation average of normalised optical chirality $\ldocoabar$. \texttt{pol\_type} is the polarisation type, \texttt{r} is a matrix containing the cartesian coordinate of the grid points. \texttt{TIJ} is the scatterer-centred \tmatrix\ of the structure.
\item \subroutine{calcTrace(TRANSA, TRANSB, A, B, {\color{\out}tr})}\\
Calculates the trace of a product of two matrices, \texttt{op(A)*op(B)}. The input characters \texttt{TRANSA} and \texttt{TRANSB} determine the operation \texttt{op}, following the convention of \BLAS' \texttt{gemm}. Specifically, \texttt{op = 'N'} corresponds to $\mathtt{op(A) = A}$ (no operation), whereas \texttt{op = 'C'} corresponds to $\mathtt{op(A) = A^\dagger}$. 
\end{itemize}
\begin{itemize}
\item  \function{RotMatX(ang) result({\color{\out}rotMat})}\\
Calculates a rotation matrix along the x axis using input argument angle(\texttt{ang}).

\item  \function{RotMatY(ang) result({\color{\out}rotMat})}\\
Calculates a rotation matrix along the y axis using input argument angle(\texttt{ang}).

\item  \function{RotMatZ(ang) result({\color{\out}rotMat})}\\
Calculates a rotation matrix along the z axis using input argument angle(\texttt{ang}).

\item  \function{rotZYZmat(angles) result({\color{\out}mat})}\\
Calculates rotation matrix \texttt{mat} for ZY'Z', using the Euler \texttt{angles}=($\alpha,\beta,\gamma$) 
\end{itemize}

\textbf{List of common arguments} 
\begin{itemize}
\item \texttt{acs\_int\_}: a matrix containing partial internal absorption inside each scatterer and for each shell.
\item \texttt{aJ($\text{nscat}\times l_{max}$), fJ($\text{nscat}\times l_{max}$)}: contains incident and scattering coefficients.
\item \texttt{Bfield}: contains the real and imaginary parts of the magnetic near field at the specified grid points, wavelengths, and incidence.
\item \texttt{ehost}: a vector of dielectric permittivity of the host medium at specified wavelengths. 
\item \texttt{escat\_(nscat, 4, size(wavelen))}: depending on the number of wavelengths, it is a 2D or 3D array of dielectric values for each scatterer, for each shell and wavelength.
\item \texttt{field}: contains the real and imaginary parts of the electric near field at the specified grid points, wavelengths, and incidence.
\item \texttt{geometry}: a matrix containing physical information of different scatterers such as centre, dimensions and direction. 
\item \texttt{ierr\_}: an integer value (0 or 1 or 2); 0 indicates solving was successful, 1 means there is an error in processing arguments, and 2 means an error in prestaging, staging, or solving/inverting $Ax=b$.
\item \texttt{iregt}: logical input, specifies whether vectors are regular or not.
\item \texttt{jregt}: logical input, specifies whether vectors are regular or not.
\item \texttt{mnq\_}: an optional logical argument which is false by default, but if true will change the indexing convention from (q,n,m) to (m,n,q), which is used to make the z-axial VTACs block-diagonal. Note that index $q$ corresponds to $s$ in this user guide.
\item \texttt{ncut}: a vector in the  form [$n_1$, $n_2$, tol], which contains the values corresponding to the keyword "\keywd{MultipoleCutoff}". Default values: $[8, 8, -8]$.
\item \texttt{nmax2\_}: an integer value equals to ncut(2).
\item \texttt{noRTR\_}: an optional input with logical value \texttt{.true.} or \texttt{.false.} for the keyword \keywd{DisableRTR}. Default: \texttt{.false.}.
\item \texttt{nselect\_}: an optional input matrix which includes information about multipole selection for different scatterers.
\item \texttt{oa\_ldoc} ($\text{npts}\times 4\times \text{nwavelen}$): contains the orientation averaged value of $\ldocbar$ at different grid points and wavelengths.
\item \texttt{orAvextE\_int}($\text{npts}\times \text{nwavelen}$): contains the orientation averaged electric field intensity values at different grid points and wavelengths.
\item \texttt{p\_label}: a matrix determining the position of each grid point, whether it is inside the surrounding medium or particles, and in which layer. 
 \item \texttt{rev\_}: an optional logical input which is false by default; triggers the reverse of balancing -- "unbalancing".
\item \texttt{right\_}: a logical input. According to (Eqs.~\ref{eqn:TjiMatrixFactored1}, \ref{eqn:TjiMatrixFactored2}) there are two ways for obtaining the \texttt{TIJ} matrix. This argument determines whether the product is taken from the left or from the right. 
\item \texttt{rtr}: a logical input that is the reverse of \texttt{noRTR\_}.
\item \texttt{scatMiet}(nscat): a logical vector with \texttt{.true.} and \texttt{.false.} values, determining whether a scatterer is spherical or not.
\item \texttt{scatXYZ}(3,nscat): a matrix containing the cartesian coordinates (in lab frame) of the particle's centre.
\item \texttt{scatXYZR}(4,nscat): a matrix containing the cartesian coordinates (in lab frame) and the radius of the smallest circumscribed sphere of each particle.
\item \texttt{scheme, scheme\_}: an integer value specifying the selected scheme.
\item \texttt{sig\_}: a matrix containing cross-sections (Extinction, Scattering, Absorption) for different polarisation(s), wavelength(s), and incidence(s). 
\item \texttt{sig\_abs\_}($4\times \text{nscat}\times 4\times \text{nwavelen}\times \text{nfi}$): a 5D array containing absorption cross-sections inside each shell for each scatterer for 4 Jones vectors, different wavelengths and different incidence directions.
\item \texttt{sig\_oa\_}($6\times \text{n}\times \text{nwavelen}$): a matrix consisting of orientation-averaged cross-sections and CD at different wavelengtha. The first column gives the values for $n_{max}$ and other columns contain values for different value of $n=1, \dots, \text{ncut(2)}$.   
\item \texttt{tfiles\_}: a matrix of character type, includes the \tmatrix\ filename and filepath for non-spherical scatterers.
\item \texttt{tol\_, rtol\_}: a real value \texttt{rtol\_G = $10^{\textbf{ncut(3)}}$}. 
\item \texttt{N\_OC}: contains $\ldocbar$ at the specified grid points, wavelengths, and incidence.
\item \texttt{verb\_}: an integer variable containing the verbosity value ($\in [0,1,2,3]$) (the default value is 1). 
\item \texttt{wavelen}: a vector of specified wavelength(s).
\end{itemize}
\subsubsection*{\module{miet} module}
This module contains routines for calculating one-body \tmatrices\ (currently limited to spherical scatterers, using Mie theory).
\begin{itemize}
\item \subroutine{calcMieTMat(x, s, zeropad\_, {\color{\out}tmat})}\\
Calculates the diagonal \tmatrix\ of a spherical scatterer for a given size parameter \texttt{$x=kR$}, relative refractive indices ($s = k_{in}/k_{out}$); \texttt{zeropad\_=nmax} maximum value of the multipole index inferred from \texttt{tmat}'s dimensions. 

\item \subroutine{calcMieCoeffs(x, s, {\color{\out}gammas, deltas})}\\
Calculates the Mie coefficients for a spherical scatterer as defined by equations H.46 and H.47 of Ref.~\citenum{LeRuE08}.
The coefficients are interpreted as magnetic and electric susceptibilities ($\Gamma_n$ and $\Delta_n$, respectively) of the scattered field. Note the relation to standard Mie coefficients\cite{MishchenkoTL02}: $a_n = -\Delta_n$ and $b_n = -\Gamma_n$.

\item \subroutine{calcCoatMieCoeffs(x, s, {\color{\out}gammas, deltas})} \\
Calculates the Mie coefficients for a coated sphere based on the equations H.110 and H.113 of Ref.~\citenum{LeRuE08}. 

\item \subroutine{calcStoutCoeffs(x, rri, nmax, {\color{\out}Cn, Dn})}\\
Calculates the \texttt{Cn,Dn} coefficients as defined by equation (50) in Stout\cite{StoutAL02}. These coefficients are used to calculate absorption cross-sections. \texttt{rri} is the relative refractive index, \texttt{nmax} is the maximum value of the multipole index. 

\item \subroutine{calcMieIntCoeffs(a, k, scaCoeffs, {\color{\out}intCoeffsReg, intCoeffsIrr, csAbs)}}\\
Calculates the regular and irregular VSWF coefficients for the field inside each concentric region of a (layered) Mie scatterer. The formulae are based on Eqs.~H.117--H.123 of Ref.~\citenum{LeRuE08}.
\texttt{a}, \texttt{k}, and \texttt{scaCoeffs}  are vectors of the radius of the concentric interfaces, relative refractive index, and scattered field coefficients for the host medium, respectively.
\texttt{intCoeffsReg} and \texttt{intCoeffsIrr} are matrices of regular and irregular field coefficients for each concentric region inside the scatterers and \texttt{csAbs} contains the partial absorptions calculated using equation (29) in Mackowski\cite{Mackowski:1990aa}.
\end{itemize} 

\subsubsection*{\module{swav} module} 
This module contains routines for calculating and transforming scalar (SSWs) and vector spherical waves (VSWFs). It depends on \library{Amos (toms644.f)} to calculate spherical Bessel and Hankel functions using recurrence. In order to limit redundancy, parameter definitions are renewed only where they are changed.    
  
\begin{itemize}
\item \subroutine{calcVTACs(r0, k, regt, {\color{\out}vtacs})} \\
Calculates the irregular (if \texttt{regt=.false.}) or the regular (if \texttt{regt=.true.}) vector translation-addition coefficients for a given \texttt{kr0}. \\ 
\texttt{r0} is a relative position vector, \texttt{k} is the wavenumber, \texttt{regt} is a logical argument which determines the type: regular or irregular, and \texttt{vtacs(1:2*pmax,1:2*pmax)} is the input/output array.     

\item \subroutine{calcSTACs(r0, k, pmax, regt, {\color{\out}scoeff)}} \\
Calculates the scalar translation-addition coefficients.($\alpha_{nu,mu;n,m}$ or $\beta_{nu,mu;n,m}$). The output corresponds to the scalar translation-addition coefficients \texttt{$\alpha$}(irregular, for \texttt{regt=.false.}) or \texttt{$\beta$}(regular, for \texttt{regt=.true.}).\\
\texttt{pmax} is a maximal composite index and \texttt{scoeff(0:pmax,0:pmax)} is the coefficients matrix.

\item \subroutine{calcVTACsAxial(r0, k, pmax, regt, flip, mqn\_, {\color{\out}vtacs})} \\
Calculates the irregular (if \texttt{regt=.false.}) or the regular (if \texttt{regt=.true.}) vector translation-addition coefficients for a given \texttt{kr0}, along the z-axis.\\
\texttt{r0} is the z-axial displacement distance, flip is a logical argument, \texttt{mqn\_} is a logical argument for changing from \texttt{qnm} to \texttt{mqn} indexing, and   
\texttt{vtacs(1:2*pmax,1:2*pmax)} is the matrix of coefficients.

\item \subroutine{calcSTACsAxial(r0, k, pmax, regt, flip, {\color{\out}stacs})}\\
Calculates the normalised scalar translation-addition coefficients along the z-axis for a given \texttt{kr0}.\\
\texttt{r0} is a displacement distance and \texttt{stacs(0:pmax,0:pmax)} is the coefficients matrix corresponding to $\alpha$ (irregular, for \texttt{regt=.false.}) or $\beta$.

\item \subroutine{calcVSWs(r, k, pmax, regt, cart, {\color{\out}waves, wavesB})} \\
Calculates (at \texttt{r}) the normalised vector spherical waves, $M_{nm}$ and $N_{nm}$ for evaluation of electric and magnetic fields \\
\texttt{r(3)} is the cartesian coordinate of a point in 3D; \texttt{cart} is a logical argument which triggers conversion to cartesian coordinates;
\texttt{waves(2*pmax,3)} contains elements ($M_{nm}$ and $N_{nm}$) of the abstract column vector defined in Eq. B1 of Ref.~\citenum{StoutAL02} and \texttt{wavesB(2*pmax,3)} is similar to \texttt{waves}, only swapping the position of $M_{nm}$ and $N_{nm}$ and multiplying by $-ik$ for calculation of the magnetic field.      

\item \subroutine{calcSSWs(xyz, k, pmax, regt, {\color{\out}psi})}\\
Calculates (at \texttt{xyz}) the scalar spherical waves \texttt{$\psi_{nm}$}. \\
\texttt{xyz} is the cartesian coordinates of a point in 3D; \texttt{psi(0:pmax)} contains elements of the spherical waves \texttt{$\psi_{nm}$} as defined by equation 13a in Chew\cite{Chew92}.

\item \subroutine{calcJCoeffsPW(ipwE0, kVec, xyz, {\color{\out}ipwCoeffsJ})}\\
Translates the supplied \texttt{ipwCoeffs} coefficients to different centres for an incident plane wave. \\
\texttt{ipwE0(3)} is the incident plane wave's amplitude vector, \texttt{kVec(3)} is the incident wave vector, \texttt{xyz(3,nscat)} is a matrix containing the centre of different scatterers, and \texttt{ipwCoeffsJ}($\text{nscat} \times \text{lmax}$) is a vector containing the translated incident plane wave coefficients to the centre of different scatterers (according to equation 38 of Ref.~\citenum{StoutAL02}). 

\item \subroutine{calcCoeffsPW(ipwE0, ipwDirn, {\color{\out}ipwCoeffs})}\\ 
Calculates the coefficients for expressing an incident plane wave in terms of vector spherical waves $M_{nm}$ and $N_{nm}$.\\
\texttt{ipwDirn}(3) is the normalised direction vector of the incident plane wave and \texttt{ipwCoeffs}(2*pmax) contains coefficients for expressing an incident plane wave in terms of vector spherical waves $M_{nm}$ and $N_{nm}$, up to a maximum $n_{max}$. Follow equations C.57-59 on p.377 of Ref.~\citenum{MishchenkoTL02}.    
 
\item \subroutine{offsetCoeffsPW(a, kVec, xyzr, {\color{\out}aJ})}\\
Translates the VSWF coefficients (\texttt{a}) of an incident plane wave (centred at the origin) to another origin.\\
\texttt{a}($l_{max}$) contains coefficients for a regular VSWF expansion centred at the origin for an incident plane wave, \texttt{xyzr} includes centres of different scatterers, and \texttt{aJ} contains scatterer centred coefficients.    

\item \subroutine{calcWignerBigD(angles, pmax, {\color{\out}bigD})}\\
Calculates the Wigner D-functions ($D^s_{m,n}(\alpha,\beta,\gamma)$).\\
\texttt{angles}(3) includes ($\alpha,\beta,\gamma$) in radians and \texttt{bigD}(pmax,pmax) contains Wigner D-coefficients.  
    
\item \subroutine{calcWignerLittled(theta, pmax, {\color{\out}d})}\\
Calculates the Wigner d-functions ($d^s_{m,n}(\theta)$).\\
\texttt{theta} is angle in radians and \texttt{d}(0:pmax,0:pmax) are values for $d^s_{m,n}$ in block diagonal matrix form. 

\item \subroutine{calcWignerd0andMore(x, pmax, {\color{\out}d, pi, tau})}\\
Calculates the Wigner d-functions for $n=0$ and also computes the derivative functions for optional outputs \texttt{pi} and \texttt{tau}.\\
\texttt{x} is $\cos(\theta)$, \texttt{d}(0:pmax), \texttt{pi}(0:pmax), \texttt{tau}(0:pmax) contain values for $d^s_{m,0}$, $\pi_{m,s}$, and $\tau_{m,s}$ respectively.     

\item \subroutine{calcRiccatiBessels(z, nmax, regt, {\color{\out}f, df})}\\
Calculates the Riccati-Bessel functions $\psi_n$ (if \texttt{regt=.true.}) or $\xi_n$ (\texttt{regt=.false.}), and their derivatives, for $n=1,\dots,\nmax$.\\
\texttt{z} is a scalar complex argument, \texttt{f}(1:nmax) is a matrix containing Riccati-Bessel functions $\psi_n(z)=z*j_n(z)$ or $\xi_n(z)=z*h_n(z)$ for $n=1,\dots,\nmax$, and \texttt{df}(1:nmax) are the corresponding derivatives of \texttt{f}.  

\item \subroutine{calcSphBessels(z, nmax, regt, {\color{\out}bes})}\\
A wrapper routine for computing spherical Bessel/Hankel functions of the first kind for a complex argument \texttt{z}.\\
\texttt{bes(0:nmax)} contain Bessel ($J_{n+1/2}$) or Hankel ($H_{n+1/2}$) function (of $1^{st}$ kind) values for $n=0,\dots,\nmax$ for a complex argument \texttt{z}.  

\item \subroutine{xyz2rtp(xyz, {\color{\out}rtp, cth})}\\
Transforms the cartesian coordinates \texttt{(x,y,z)} of a point in 3D space to spherical polar coordinates \texttt{($r,\theta,\phi$)}\\
\texttt{xyz(3)} is a vector of cartesian coordinates, \texttt{rtp(3)} is a vector of spherical polar coordinates, and \texttt{cth} is cos$(\theta)$. 

\item \subroutine{rtp2xyz(rtp, {\color{\out}xyz})}\\
The inverse of \texttt{xyz2rtp}. Transforms the spherical polar coordinates \texttt{($r,\theta,\phi$)} of a point in 3D space to cartesian coordinates $(x,y,z)$.

\item \subroutine{calcVTrtp2xyz(rtp, {\color{\out}transform})}\\
Calculates the matrix of transformation from a vector in spherical coordinates to a vector in cartesian coordinates at point \texttt{($r,\theta,\phi$)} (in spherical polar coordinates).

\item \subroutine{calcVTxyz2rtp(rtp, {\color{\out}transform})}\\
The inverse of \texttt{calcVTrtp2xyz}. Calculates the matrix of transformation from a vector in cartesian coordinates to a vector in spherical coordinates at point \texttt{($r,\theta,\phi$)} (in spherical polar coordinates).\\
\item  \subroutine{calcAbsMat(Xi, ro, {\color{\out}mat})}\\
Calculates the absorption matrix $\Gamma_j = \mathtt{mat(l_{max},l_{max})}$ for the input arguments \texttt{Xi} and \texttt{ro} (Eq.~(49) of Ref.~\citenum{StoutAL02}). $\Gamma_j$ is used in the evaluation of the orientation-averaged absorption cross-section inside each particle. 
\item  \subroutine{calcLamMat(Xi, ro, {\color{\out}mat})}\\
Calculates the "Lambda" matrix $\Lambda_j  = \mathtt{mat(l_{max},l_{max})}$ for the input arguments \texttt{Xi} and \texttt{ro} (Eq.~(53) of Ref.~\citenum{StoutAL02}). $\Lambda_j$ is used in the evaluation of the orientation-averaged internal electric field inside homogeneous spheres.
\item \subroutine{nm2p(n, m, {\color{\out}l})}\\
Calculates a generalised index \texttt{l=n(n+1)+m}, for a unique \texttt{(n,m)}, (Vector spherical harmonics are spanned by two indices: $n$ and $m$, such that $0 \leq n \leq n_{max}$ and $-n \leq m \leq n$).\\
\texttt{n,m,l} are integers. 

\item \subroutine{p2nm(p, {\color{\out}n, m})}\\
Calculates unique \texttt{(n,m)} from a given composite index \texttt{p}.\\
\texttt{p} is a real value.

\item \subroutine{nm2pv2(n, m, {\color{\out}p})}\\
Some recurrences are defined only for $m \geq 0$, in which case we shall use a second version of the composite index $p_{v2}=n(n+1)/2+m$. 

\item \subroutine{testPmax(name, pmax, {\color{\out}nmax})}\\
Tests \texttt{pmax} for commensurability, i.e. is $p_{max}==n_{max}(n_{max}+2)$ and $n_{max}=m_{max}$? If not, then the program will be stopped.
\end{itemize}
\subsubsection*{\module{sphmsv} module}
This module contains routines for calculating Stokes incident vector, Stokes phase matrix and scattering matrix for an input \tmatrix. The formulae are based on Mishchenko\cite{MishchenkoTL02}.  
\begin{itemize}
\item \subroutine{calcStokesIncVec(ehost\_, ipwDirn\_, ipwAmpl\_, verb\_, {\color{\out}Stokes\_Vec})}\\
Calculates the Stokes incident vector \texttt{Stokes\_Vec}.

\item \subroutine{calcStokesPhaseMat(SMat, verb\_, {\color{\out}Z})}\\
Calculates the Stokes phase matrix for the specified incident and scattered angles. \texttt{SMat(2,2)} and \texttt{Z(4,4)} are the scattering and Stokes phase matrices which follow Eqs. (5.11-14) and (2.106-121) of Ref.~\citenum{MishchenkoTL02}. 

\item \subroutine{calcScatMat(tmat, host\_K, spwDirn\_, ipwDirn\_, verb\_, {\color{\out}SMat})}\\
Calculates the scattering matrix using the \tmatrix, for the specified incident and scattering angles.

\end{itemize} 
\subsubsection*{\module{linalg} module}
This module contains wrappers to drive LAPACK's square-matrix inversion routines and linear solvers.
\begin{itemize}
\item \subroutine{invSqrMat(trans\_, verb\_ {\color{\out}A})}\\
Calculates inverse of  a complex-valued square matrix \texttt{A(n,n)}, using the ZGETRF and ZGETRI routines in LAPACK. \texttt{A} is overwritten by inv(A) on  the output. \texttt{trans\_} is an optional logical input, in case \texttt{.true.} the routine considers transpose of A and finally returns the transpose of the inverted matrix as the output. \texttt{verb\_}: an optional input of the verbosity value. \\

\item \subroutine{solLinSys(isol\_, verb\_, {\color{\out}A, X})}\\
Solves a complex-valued linear system of equations $Ax=b$, where \texttt{A}(n,n) is a square matrix, \texttt{b}(n) is a known vector, and \texttt{x}(n) is the vector to be determined. Depending on the value \texttt{isol\_}, calls \texttt{solLinSysV} or \texttt{solLinSysVX}. Both \texttt{A} and \texttt{X} are overwritten on output.

\item \subroutine{solLinSysV(verb\_, {\color{\out}A, X})}\\
For solving a linear system, uses LAPACK's "simple" driver ZGESV. 
\item \subroutine{solLinSysVX(verb\_, {\color{\out}A, X})}\\
For solving a linear system, uses LAPACK's "simple" driver ZGESVX. 

\end{itemize}
\subsubsection*{\module{eps} module}
This module contains wavelength-dependent dielectric functions \texttt{epsXX(lambda)} for various materials including Au, Ag, Al, Cr, Pd, Pt, Si, and Water). 
\begin{itemize}
\item \subroutine{interp1( x1, y1, x2, {\color{\out}y2} )}\\
Calculates the interpolated data \texttt{y2} using the input values \texttt{x1,y1} at the points \texttt{x2}.
\end{itemize}
\begin{itemize}
\item  \function{epsAu(wavelength) result({\color{\out}eps})}\\
Returns the wavelength-dependent relative dielectric function of gold.
This function uses the analytical expression given in Eq.~(E.2) of Ref.~\citenum{LeRuE08}.

\item  \function{epsAg(wavelength) result({\color{\out}eps})}\\
Returns the wavelength-dependent relative dielectric function of silver.
This function uses the analytical expression given in Eq.~(E.1) of Ref.~\citenum{LeRuE08}.

\item  \function{epsPt(wavelength) result({\color{\out}eps})}\\
Returns the wavelength-dependent relative dielectric function
of a Lorentz-Drude metal, with the parameters for Pt\cite{rakic1998optical}.
\item  \function{epsPd(wavelength)
result({\color{\out}eps})}\\
Returns the wavelength-dependent relative dielectric function
of a Lorentz-Drude metal, with the parameters for Pd\cite{rakic1998optical}.

\item  \function{epsSi(wavelength) 
result({\color{\out}eps})}\\
Returns the wavelength-dependent relative dielectric function of Silicon in the range 206.6 nm to 1200.0 nm interpolated from \citep{aspnes1983dielectric}.

\item  \function{epsAl(wavelength) 
result({\color{\out}eps})}\\
Returns the wavelength-dependent relative dielectric function
of Aluminum in the range 103.32 nm to 2755.2 nm \cite{rakic1995algorithm}.
\item  \function{epsCr(wavelength) 
result({\color{\out}eps})}\\
Returns the wavelength-dependent relative dielectric function of Aluminum in the range 100.8 nm to 31 $\mu m$, from the tabulated data in Ref.~\citenum{palik1998handbook} pages: 382-385. 

\item  \function{epsWater(wavelength) 
result({\color{\out}eps})}\\
Returns the wavelength-dependent relative dielectric function of Water at temperature $20^o\text{C}$ in the range 200 nm to 3000 nm \cite{daimon2007measurement}.

\item  \function{epsDrude(wavelength, eps\_infty, lambda\_p, mu\_p) result({\color{\out}eps})}\\
Returns the wavelength-dependent relative dielectric function of a Drude metal. The analytical expression is given in Eq.~(3.2) of Ref.~\citenum{LeRuE08}.
 
\end{itemize}
\subsubsection*{\module{HDFfive} module}
This module contains subroutines for reading and writing data in HDF5 format. 
\begin{itemize}
\item\subroutine{h5\_crtgrp(filename\_, main\_grpname, subgrpsname)}\\
This subroutine creates subgroups in an existing group.
\item\subroutine{h5\_wrtvec2file(filename\_, groupname, dsetname, dset\_data)}\\
This subroutine writes vector data in a dataset in an existing group.
\item\subroutine{h5\_wrt2file(filename\_, groupname, dsetname, dset\_data)}\\
This subroutine writes data in a dataset in an existing group.
\item\subroutine{h5\_wrt\_attr(attribute, dataset\_id)}\\
This subroutine adds an attribute to an existing dataset, typically a brief explanation about the contents of the dataset.
\end{itemize}
%